\documentclass[lettersize,journal]{IEEEtran}
\IEEEoverridecommandlockouts
\usepackage{cite}
\usepackage{amsmath,amssymb,amsfonts}
\usepackage{algorithmic}
\usepackage{graphicx}
\usepackage{textcomp}
\usepackage{xcolor}
\usepackage{caption}
\usepackage{subcaption}
\usepackage{wrapfig}
\def\BibTeX{{\rm B\kern-.05em{\sc i\kern-.025em b}\kern-.08em
    T\kern-.1667em\lower.7ex\hbox{E}\kern-.125emX}}
\begin{document}

\title{Faster Region-Based CNN Spectrum Sensing and Signal Identification in Cluttered RF Environments}

\author{Todd Morehouse, Charles Montes, and Ruolin Zhou\\
Department of Electrical and Computer Engineering\\ University of Massachusetts, Dartmouth, MA
}
\maketitle

\begin{abstract}
In this paper, we optimize a faster region-based convolutional neural network (FRCNN) for 1-dimensional (1D) signal processing and electromagnetic spectrum sensing. We target a cluttered radio frequency (RF) environment, where multiple RF transmission can be present at various frequencies with different bandwidths. The challenge is to accurately and quickly detect and localize each signal with minimal prior information of the signal within a band of interest. As the number of wireless devices grow, and devices become more complex from advances such as software defined radio (SDR), this task becomes increasingly difficult. It is important for sensing devices to keep up with this change, to ensure optimal spectrum usage, to monitor traffic over-the-air for security concerns, and for identifying devices in electronic warfare. Machine learning object detection has shown to be effective for spectrum sensing, however current techniques can be slow and use excessive resources. FRCNN has been applied to perform spectrum sensing using 2D spectrograms, however is unable to be applied directly to 1D signals. We optimize FRCNN to handle 1D signals, including fast Fourier transform (FFT) for spectrum sensing. Our results show that our method has better localization performance, and is faster than the 2D equivalent. Additionally, we show a use case where the modulation type of multiple uncooperative transmissions is identified. Finally, we prove our method generalizes to real world scenarios, by testing it over-the-air using SDR.
\end{abstract}

\section{Introduction}
\label{sec:intro}

Wireless communications continue to become more prolific and more complex due to cheapening hardware and advances in fields like machine learning (ML) and software defined radio (SDR). Sensing becomes an increasingly difficult task as frequency congestion increases and device use increasingly complex patterns. Machine learning has improved a device's ability to adapt, with technologies such as online learning allowing new patterns, behaviors, and capabilities, to be realized in the field. Additionally, wireless devices continue to be deployed in increasingly diverse environments, including very urban and very remote areas. These devices experience very diverse wireless channels in congestion levels, noise levels, and multipath effects. These changes and effects all impact spectrum sensing, the process of identifying signals within a wide spectrum band. Spectrum sensing is vital for allocating spectrum resources for communications, and for organizations to monitor activities. Adaptive radios must ensure to use an available portion of the spectrum, and do not interfere with devices that have higher priority. To achieve this, devices must accurately locate any other transmissions in a band. Government, military, and commercial sectors may wish to monitor communications for security threats over RF, or improper use of spectrum. Electronic warfare may wish to identify adversary communications or interference. The ability to accurately locate each transmission present is paramount for each of these tasks.

Popular traditional spectrum sensing techniques include energy-based and eigenvalue-based detection methods, such as MED, and AGM, which require prior knowledge of noise power \cite{zhao_eigenvalues-based_2021}. Newer eigenvalue-based methods have addressed the drawbacks, but are still inflexible in domains with dynamic users \cite{zhao_eigenvalues-based_2021, liu_maximum_2019}. Deep learning methods often employ techniques, such as convolutional neural networks (CNNs), that do not require prior knowledge of the signal-to-noise ratio (SNR) conditions \cite{xie_deep_2020}. APASS, implemented by Xie et al [\citenum{xie_activity_2019}] is a CNN method to identify if a primary user is active on a channel, and outperforms traditional detectors. More recently, object detection and segmentation methods have been used to localize signals in multi-user environment. In O'Shea et al \cite{oshea_learning_2017}, the authors showed that you-only-look-once (YOLO) can be used to find signals present in a spectrogram. In Prasad et al \cite{prasad_deep_2020, prasad_downscaled_2020} the authors used faster region-based CNN (FRCNN) to jointly detect signals within the spectrogram image, and classify detections as Bluetooth, WiFi, or microwave oven interference. They were able to achieve a mean average precision (mAP) of 0.713 under the SNR interval between 15dB and 50dB. They tested their system over-the-air (OTA), but were only able to achieve a mAP of 0.125. In Vagollari et al \cite{vagollari_joint_2021}, the authors used YOLO and FRCNN to localize signals within a spectrogram, and classify them by their modulation type. They were able to achieve an mAP of 87\% and a generalized intersection over union (gIoU) of 90\%. They found that FRCNN outperforms YOLO for spectrum sensing. Other authors investigated object detection on frequency contents only. In Ghanney and Ajib \cite{ghanney_radio_2020}, the authors locate RF interference within a FFT generated plot. In order to process the plot with YOLO, they save images of the plots. Their results achieve an mAP of 0.81 at an IoU of 0.5. In Kayraklik et al \cite{kayraklik_application_2022}, the authors generate power spectral density (PSD) plots from baseband signals collected OTA. They save their PSD plots as images so that they can be processed by You Only Learn One Representation (YOLOR) and Detectron2.

\subsection{Challenges in Current Research}
Our literature review shows that current implementations are limited to images by their object detection baselines. In this paper, we show that by optimizing FRCNN to process 1D FFTs directly, we can significantly reduce computational cost. In signal processing, especially wireless signals, data is often one dimensional. In the spectrum sensing case, if time information is not required, then the process can be performed with only a single dimension. In order to create images, additional information needs to be added or extracted. Using spectrograms can organize time information as well as spectral information, however, this requires significantly more collected data, and the additional data requires more computation to perform detection. In one paper, the authors render an FFT as a plot to generate an image \cite{ghanney_radio_2020}, but this introduces more data without new information. These approaches introduce two major issues: an introduction of a second dimension increases search space and computational complexity, and in the FFT plot case, adding sources of error due to the limitations of rendering an image. Object detection in a 2D case requires exponentially higher computation than for a 1D case. We show a cost comparison between the two techniques in Sec. \ref{sec:1d_2d_cost}. Plotting an FFT to generate an image is an effective way to utilize standard FRCNN to perform 1D spectrum sensing, however it is costly in both computation and performance. The increased dimension will increase data size, without adding any additional information. Furthermore, when the plot is rendered, points and lines that are close together overlap, blurring them together. This can be seen in Fig. \ref{fig:1d_example}, where we plot an FFT. The resolution in parts with high variance is lost, resulting in a solid shape.

\subsection{Cost of 2D and 1D Spectrum Sensing}
\label{sec:1d_2d_cost}
In this example, we show the difference in data size and acquisition time between a 2D and 1D spectrum sensing scenario. First, consider a receiver operating at a 200kHz baseband sample rate, and using an FFT size of 1024. Thus, in order to create one vector of size 1024 for spectrum sensing, we must receive samples for $t=1024/200000=5.12ms$. If we wish to generate a square spectrogram with the same FFT size, then we must collect $1024^2$ samples, or approximately $1.05$ MS. Thus, we would need to record for $5.24$s. Thus, to achieve the same frequency sensing resolution, we would need to record 1024x longer, and 1024x more data. We can reduce this cost by reducing the FFT size to something smaller, like 128. This still results in recording for 128x longer and 128x more data, while having a lower frequency resolution. Thus, we can see that a 1D scenario can be significantly more favorable for performing spectrum sensing.

\subsection{Our Approach}
We identify and localize RF signals in frequency domain using 1D FRCNN. Additionally, we consider a dynamic environment, where no centralized channel allocation is present. This allows wireless devices to transmit at any frequency and bandwidth, requiring a more sophisticated detection. The FRCNN algorithm is a modification to CNNs, allowing object detection and segmentation. It is applied in this project to detect and isolate received signals. The standard implementations of FRCNN are intended for processing images, and are not fit for processing received baseband samples directly. Therefor, we optimized FRCNN for the 1D signal case by modifying the ML architecture and optimizing anchor box proposals. In a use case, we show how each detected signal can be separated, and classified by its modulation type. This can be extended to further applications in classification or signal processing, such as performing security related checks, or locating a radio to communicate with. We test our system over-the-air (OTA), using two USRP N2901 software defined radios (SDRs). These devices allow controlling the physical layer of wireless communications using high level software, and provide an easy platform to integrate machine learning with communications. We used these radios to broadcast test signals, which were received on a separate radio, and processed with our spectrum sensing algorithm.

\subsection{Novelty and Contribution of our Research}
Our work optimizes FRCNN for 1D signals, greatly reducing the computational complexity, and improving the performance in spectrum sensing for detecting and locating signals. The contributions of our research are as follows
\begin{enumerate}
    \item Redesigning the ML architecture of FRCNN for 1D signals
    \item Optimization of the feature extraction layers for spectrum sensing
    \item Application of 1D FRCNN for spectrum sensing in cluttered and unknown RF environments
\end{enumerate}
The ML architecture of FRCNN consists of operations designed for working with 2D features from images. These operations had to be redesigned to process 1D signals. Additionally, several stages of FRCNN involve preprocessing the data, including generating a dataset to train the region proposal network, pooling features for the classifier, and selecting detections after classification. These stages need to handle 1D inputs and features. The feature extraction layers of FRCNN are typically designed to extract relevant image features, thus, state-of-the-art networks such as RESNET should be ideal. However, signal features are not the same, and may not be analogous to image features. We first evaluate different feature extractors for the spectrum sensing application, then optimize the best extractor for performance and inference time. Finally, we show that our optimizations provide superior performance over 2D FRCNN with spectrograms as well as over a baseline energy-based detection method. Additionally, we show that our optimizations are significantly faster than 2D FRCNN. We measured the mAP, mIoU, probability of detection (Pd), and probability of false alarm (Pfa) over the SNR range of -5dB to +20dB. The mAP was measured to be 0.716, mIoU as 0.586, Pd as 0.823, and Pfa as 0.166. Directly comparing metrics to other papers is difficult, due to differences in simulated channel conditions and metric measurement methods. In Vagollari et al \cite{vagollari_joint_2021} the authors use the generalized IoU, while we use the simple IoU, and they do not provide an SNR for the measured gIoU. In Prasad et al \cite{prasad_downscaled_2020}, the authors calculate the interpolated mAP, but do not give a probability threshold, which is typical for this method, and in Ghanney and Ajib \cite{ghanney_radio_2020}, the mAP is not well defined. Therefor, we implement a 2D spectrogram spectrum sensing method using FRCNN, using techniques from these papers, to give an accurate comparison of our results. However, if we give a direct comparison between mAP, ours is comparible with Prasad et al \cite{prasad_downscaled_2020}, while being lower than Ghanney and Ajib \cite{ghanney_radio_2020}. Our OTA mAP significantly outperforms Prasad et al \cite{prasad_downscaled_2020}, where ours measured 0.826 compared to their 0.125. When comparing with our implementation of 2D FRCNN, we show significant improvement mIoU by 91\%, our primary metric, with slightly improved mAP by 4\%.

\subsection{Paper Organization}
This paper is organized into the following sections: First we describe our optimized FRCNN model in Sec. \ref{sec:frcnn}. In Sec. \ref{sec:rf_environment}, we describe the RF environment that we apply our work to. In Sec. \ref{sec:methodologies} we describe each spectrum sensing methodology we test in this paper, including our approach. In Sec. \ref{sec:metrics} we discuss how we measure performance. In Sec. \ref{sec:architecture_analysis} we analyze different FRCNN parameters and their effect on performance, and in Sec. \ref{sec:metrics} we go into detail on how we test each methodology. In Sec. \ref{sec:amc}, we show a use-case of our 1D FRCNN with Automatic Modulation Classification, and we show our test results for simulation and over-the-air in Sec. \ref{sec:results}. Finally we conclude on the research in Sec. \ref{sec:conclusion}. 

\section{Optimizing FRCNN}
\label{sec:frcnn}
\begin{figure}[htp]
    \centering
    \includegraphics[width=0.3\textwidth]{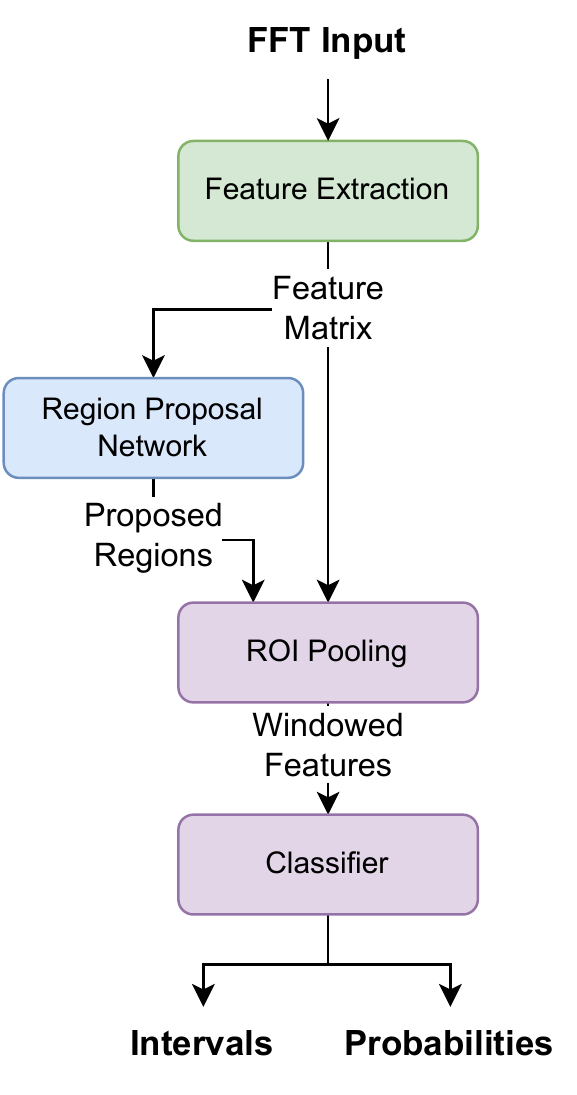}
    \caption{FRCNN Flowchart}
    \label{fig:frcnn_overview}
\end{figure}

In this section, we introduce our optimized FRCNN, design decisions that we made, and how we constructed the network.  Faster region-Based CNN \cite{ren_faster_2015} is an object detection ML model that makes heavy uses of CNNs in each stage. For spectrum sensing this allows detection of transmitters within a received signal. However, the architecture is designed for image processing. We optimize the three major components of FRCNN for 1D signal processing, including the convolutional feature extractor, region proposal network (RPN), and classifier stage, shown in order in Fig. \ref{fig:frcnn_overview}.

\subsection{Feature Extraction Base Layers}
The first part of FRCNN is feature extraction, achieved using a fully convolutional network. The goal is to have each feature map to a region in the input. For an image, this region would be a windowed subsection condensed into a single feature. Condensing intervals reduces the search space, since only the feature matrix needs to be analyzed to propose and find features. However, higher levels of feature extraction reduce the granularity, since larger sections of the input are being condensed into a single feature. Thus, a trade-off needs to be balanced between greater data reduction, feature complexity, inference time, and accuracy. Additionally, wireless signals have unique features that should be focused on. Thus, a different approach should be taken over image processing techniques. We analyzed different architectures for their feature extraction performance, as well as different levels of feature extraction. This is detailed in Sec. \ref{sec:architecture_analysis}. To ensure that the feature matrix maps directly to the input, the downscaling factor must be an integer divisor of the input size, to avoid truncation. Otherwise, this truncation would result in innaccurate mapping of features to locations in the signal space.

\subsection{Anchor Intervals and Region Proposals}
\begin{figure}[htp]
    \centering
    \includegraphics[width=0.3\textwidth]{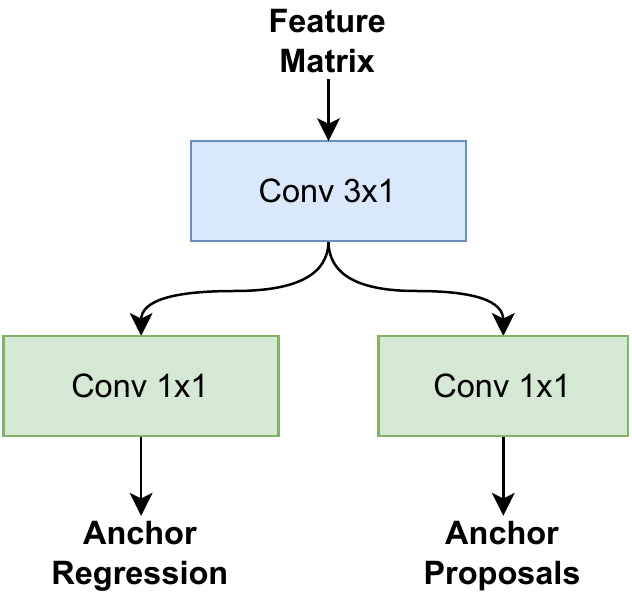}
    \caption{Convolutional layers of the RPN}
    \label{fig:rpn_architecture}
\end{figure}

Anchor intervals are an efficient way to search for objects within a dataspace, by using predefined locations and aspect ratios called anchors. The goal of the region proposal network is to identify which anchors overlap a ground truth, given an input. To achieve this, the feature matrix, which is proportional to the input size, is used. Each element of the feature matrix maps to an anchor box in the input data. The region proposal network consists of a three layer convolutional network, shown in Fig. \ref{fig:rpn_architecture}. First, a 3x1 convolutional layer serves as an intermediate layer, performing further feature extraction and setting the feature depth. The output of this is broken into two separate 1x1 convolutions, for classification and regression separately. For the classifier convolution, the number of features is set to the number of anchor boxes. The regression layer predicts intervals, therefore number of filters is twice the number of anchor boxes, allowing it to predict two points for each anchor box. The outputs are a \textit{n x k} classification matrix matrix, and an \textit{n x 2k} regression matrix, where \textit{n} represents the feature matrix size, and \textit{k} represents the number of anchors. This gives a probability and interval for each anchor box on each feature region.

\subsection{Region Pooling and Classification}
\begin{figure}[htp]
    \centering
    \includegraphics[width=0.3\textwidth]{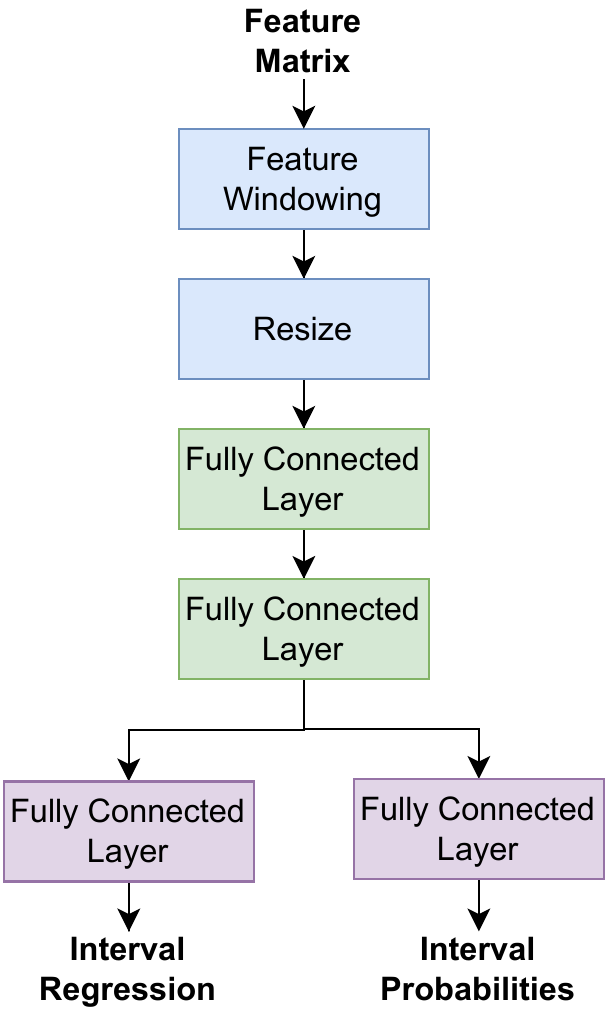}
    \caption{Deep learning layers of the classifier and ROI pooling}
    \label{fig:classifier_architecture}
\end{figure}

The region pooling and classification fine tune proposed regions from the RPN. Since the number of proposed regions will always be the same, many will be duplicates or background detections. Thus, the classifier stage must process each proposed region to evaluate if an object actually exists, and best fit a bounding interval to the object. The classifier uses regression and the feature matrix to best fit an interval to the object. The architecture for this could be seen in Fig. \ref{fig:classifier_architecture}. First, the features around each proposed region of interest are windowed, then resized to a 1x7 vector. This standard size allows it to be processed by the next step, a fully connected layer. First, the features are flattened, then processed by two fully connected layers. Finally, the result is passed to two different fully connected layers for classification and regression. The classification is either ``foreground'' or ``background'', or alternatively, a set of custom classes. The regression output is two points for each proposal and class.
The output of the classifier is a fixed number of predictions each time, however the number of objects present is not fixed. In order to output only detected objects, thresholding is applied for each output probabilities. We chose a probability threshold of 0.70. That is, any detection with a probability less than 0.70, was considered background and discarded. This allows only very confident predictions to be output.

\subsection{Training the Model}
We train the network using a two-step training process, where we alternate between training the RPN and the classifier. First, the RPN is trained on an input signal to produce proposals. Then, the RPN proposals, and the feature matrix, are used to train the classifier. These two steps are alternated until the entire model is fully trained. To achieve this, an intermediate dataset is needed to train the RPN, consisting of classification and regression targets for anchors. The downscaling factor between the input signal and feature matrix is determined by the feature extraction layers. Thus, the dimension of the feature matrix is known, and the anchor locations can be found from this. For each feature, the anchor locations are screened for overlaps with ground truths by their IOU, calculated using Eq. \ref{eq:iou}. If the IoU with any ground truth is found to be greater than 0.7, then it is considered a positive overlap. Likewise, an IoU less than 0.3 is considered a negative overlap. If the IoU is between 0.3 and 0.7, then the overlap is considered ambiguous, and not used for training. This ambiguous classification helps to ensure that the RPN is learning on confident samples. For each feature, a classification is given as ``foreground'' or ``background'', and a regression is given as the anchor box that it overlaps.

\begin{equation}
\label{eq:iou}
\text{IOU(A,B)}=\frac{|A\cap B|}{|A\cup B|}
\end{equation}

\begin{equation}
\label{eq:binary_crossentropy}
L = -\frac{1}{N}\sum_{i}^{N}{p^{*}_{i}\log{(p_{i})} + (1-p^{*}_{i})\log{(1-p_{i}})}
\end{equation}

\begin{equation}
\label{eq:categorical_crossentropy}
L = -\frac{1}{N}\sum_{i}^{N}{p^{*}_{i}\log{p_{i}}}
\end{equation}

\begin{equation}
\label{eq:smooth_l1}
L = \left \{
    \begin{array}{lr}
        \frac{1}{2}(t_{i}-t^{*}_{i})^{2} & \text{if}|t_{i} - t^{*}_{i}| < 1.0\\
        |t_{i} - t^{*}_{i}| - \frac{1}{2} & \text{otherwise}
    \end{array}
    \right\}
\end{equation}

We use separate loss functions for the RPN classification and regression outputs. The classifier uses binary crossentropy loss, shown in Eq. \ref{eq:binary_crossentropy}. The loss is calculated between the predicted probability for each proposal, $p_{i}$, and an indicator if the proposal overlaps a ground truth, $p^{*}_{i}$. The indicator is 1 for a positive overlap, and 0 for a negative overlap. The RPN regression uses smooth L1 loss, shown in Eq. \ref{eq:smooth_l1}, where $t_{i}$ represents the predicted regression target, and $t^{*}_{i}$ represents the ground truth overlap. The final classifier stage consists of separate classification and regression outputs. The regression output is trained with the same smooth L1 loss used for the RPN, shown in Eq. \ref{eq:smooth_l1}. The classifier uses categorical crossentropy, shown in Eq. \ref{eq:categorical_crossentropy}. The predicted probabilities are represented by $p_{i}$, while the true probabilities are represented by $p^{*}_{i}$. This is used since the FRCNN final classifier can have multiple classes, instead of simply ``background'' and ``foreground''.

\section{RF Testbed}
\label{sec:rf_environment}
\begin{figure}[htp]
    \centering
    \includegraphics[width=0.4\textwidth]{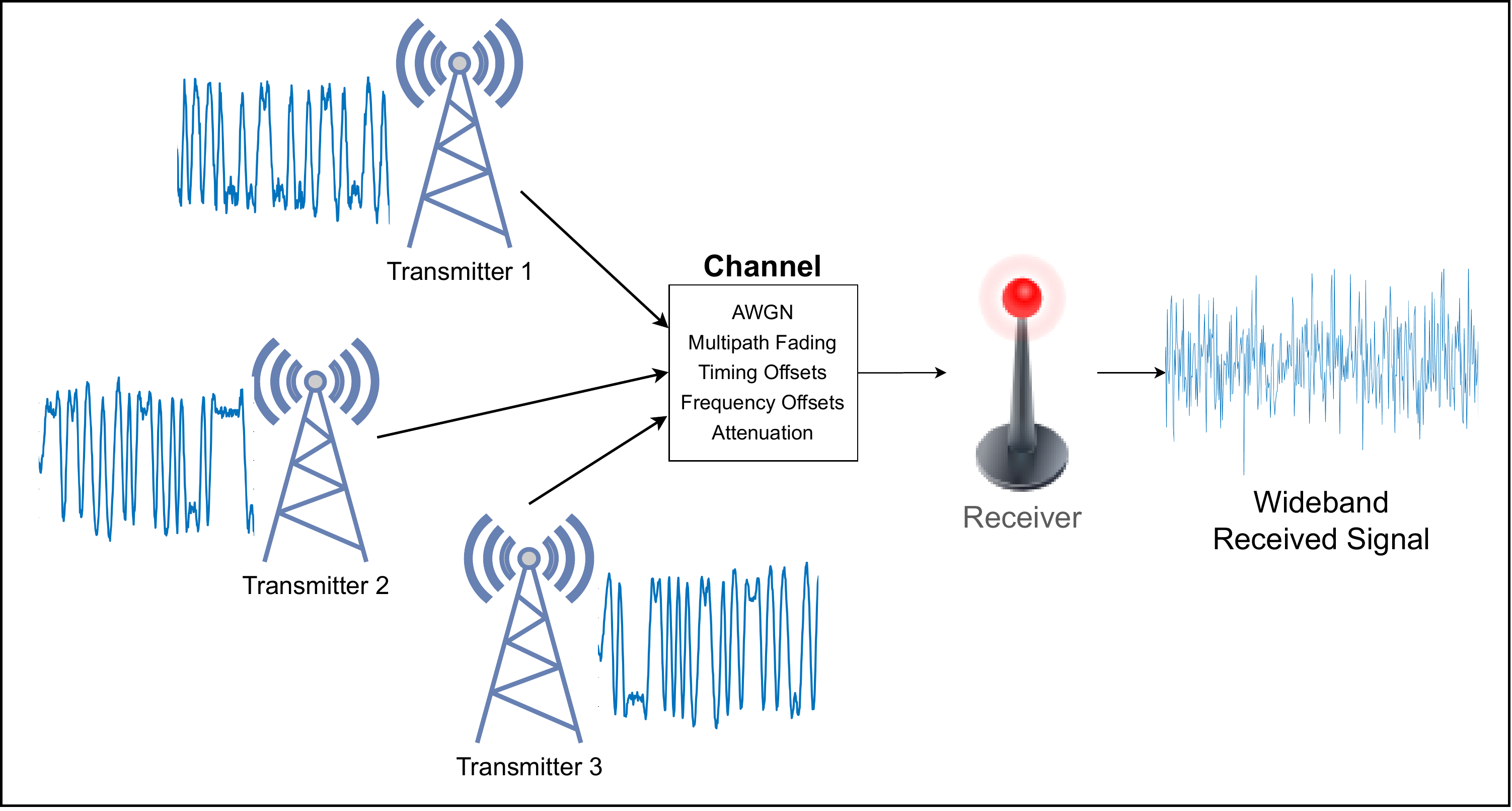}
    \caption{Illustration of a wideband receiver}
    \label{fig:rf_testbed}
\end{figure}

In this research, we focus on determining the frequency location of each transmission, from the perspective of a receiver. This is illustrated in Fig. \ref{fig:rf_testbed}, where a single receiver observes multiple transmitters. The goal of the optimized FRCNN sensing algorithm is to locate all signals present in frequency domain. In this section, we discuss the wireless environment that we design the model around, how we synthesize the environment to train and test the network, and how we construct an OTA test to evaluate the model.

\subsection{Wireless Environment}
\begin{figure}[htp]
    \centering
    \includegraphics[width=0.4\textwidth,trim={4cm 8.5cm 4cm 9cm}]{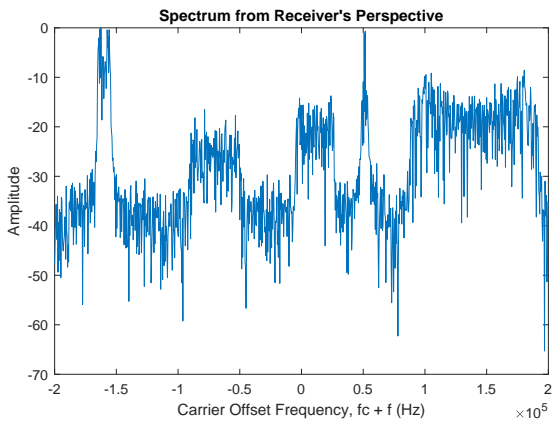}
    \caption{Receiver's Perspective of the Wireless Environment}
    \label{fig:rx_fdom_example}
\end{figure}

First, we define the wireless environment that we are operating in. We consider a spectrum of arbitrary size, where multiple transmitters are present. We observe the entire spectrum from a single receiver. Each transmission may occur at any center frequency within the band, and occupy any bandwidth. An example from a receiver's perspective is shown in Fig. \ref{fig:rx_fdom_example}, where five transmissions are present, relative to the center frequency of the receiver. This scenario represents a decentralized and independent sensor, where channel allocations may be unknown or nonexistent. A simple spectrum sensing method of identifying if a channel is occupied would lose most of the information present. For many applications, this approach will not meet requirements. Instead, a better approach is to use object detection, to detect and localize each transmission present.

\subsection{Synthesized Testbed}
Machine learning models require large amounts of diverse data for training, otherwise overfitting will occur and the model will not generalize. We elected to synthesize our training data, allowing us to generate as many samples as needed, and to generate diverse scenarios through simulated channel impairments. Diverse channel impairments allows our model to be deployed in different scenarios, without requiring retraining. Collecting data OTA is possible, but significantly more challenging, since channel effects are difficult to control and data can be difficult to label. In previous research, we synthesized our training and test data, and showed that our approach generalizes well to OTA tests \cite{gravelle_deep_2019, morehouse_baseband_2020, morehouse_incremental_2021}.

\begin{figure*}
    \centering
    \includegraphics[width=0.9\textwidth]{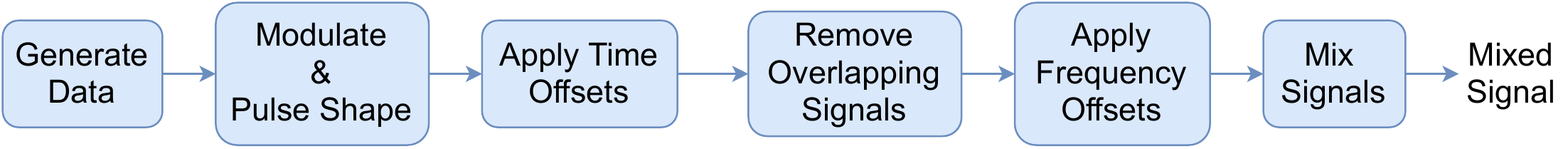}
    \caption{Process for synthesizing signals in MATLAB}
    \label{fig:matlab_synthesis}
\end{figure*}

The dataset generated must satisfy the input and output of the model shown in Fig. \ref{fig:frcnn_overview}. To generate the input, each transmitted signal is generated independently. The process for synthesizing samples could be seen in Fig. \ref{fig:matlab_synthesis}. For each signal, random binary data is generated, then modulated. The modulation type is chosen at random. Each signal is then pulse shaped using a raised cosine filter, and resampled to achieve a desired bandwidth. The signal is finally shifted by its center frequency offset in frequency domain. This process is repeated a total of five times, where each center frequency and bandwidth is chosen at random. Then, if any signals are overlapping in frequency domain, one of the signals is removed at random. This selection process results in a random number of transmitters. Finally, all of the signals are mixed together, by summing them in time domain. This process generates random environments, where the number of transmissions are random, as well as their corresponding center frequencies and bandwidths.

\begin{equation}
    \label{eq:power_norm}
    y = \frac{x}{\sqrt{\frac{1}{N}\sum{|x^2|}}}
\end{equation}

\begin{figure}[htp]
    \centering
    \includegraphics[width=0.45\textwidth]{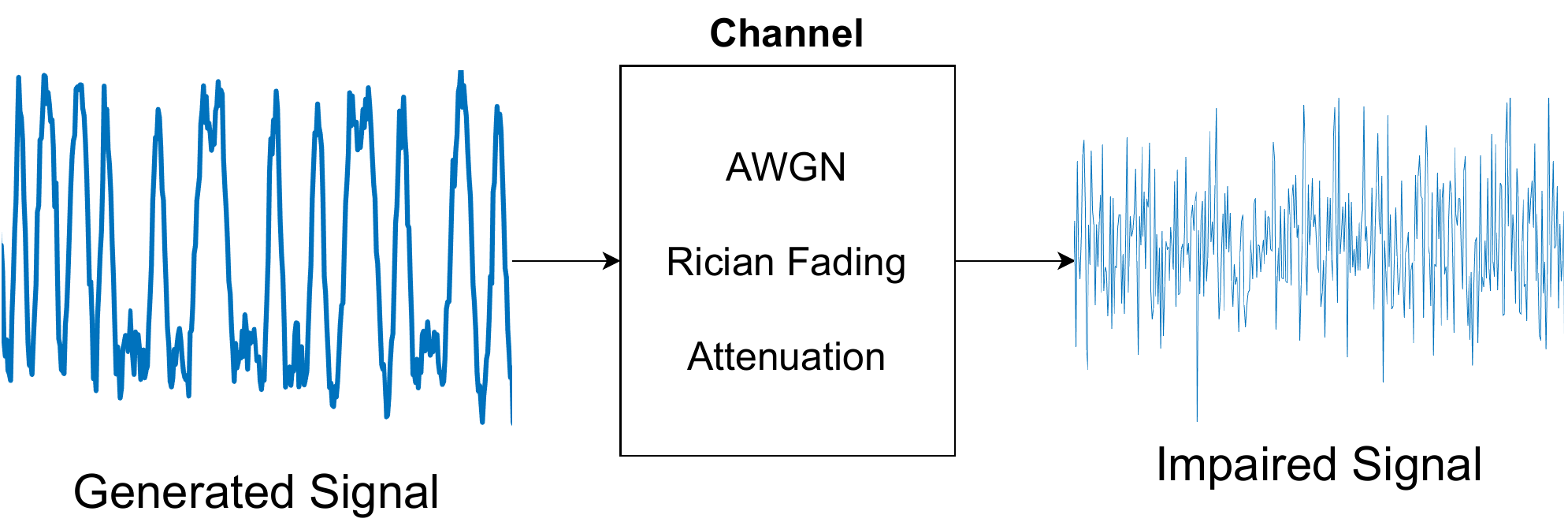}
    \caption{Channel Impairment Process for Synthesized Signals}
    \label{fig:channel_effect}
\end{figure}

Channel effects were applied to each mixed signal, as shown in Fig. \ref{fig:channel_effect}. Before AWGN could be applied, the signal's power was normalized to 1, using Eq. \ref{eq:power_norm}. We considered an indoor system, where a direct line of sight was present between all transmitters and the receiver. Thus, a Rician multipath process was used. Finally, the FFT of the signal was taken, to be used for training FRCNN, along with the frequency location of each signal present, as shown in Fig. \ref{fig:matlab_synthesis}.

This process was repeated to generate a total of 3,000 training samples.

\subsection{OTA Testbed}
An OTA testbed was designed to model Fig. \ref{fig:rf_testbed}. To create this, we used SDR, allowing us to control the transmitted waveforms, as well as to process the received waveform using machine learning. We used 2x USRP N2901, which are capable of 2x2 MIMO each. The SDRs were configured as three transmitters, and one receiver. They were placed in an indoor environment, where each transmitter had a direct line of sight path to the receiver. The radios were configured to operate at a center frequency of 5 GHz, with a sample rate of 200kHz for each channel. Each radio was calibrated to the same center frequency, to remove any center frequency offset from hardware, and thus collect more accurate metrics. To control each transmission's center frequency and bandwidth, offsets and pulse shaping were applied to the baseband signals before transmitting, following the same process for synthesizing signals. A random bitstream was generated, modulated, and pulse shaped, then resampled to a desired bandwidth. Finally, a center frequency offset was applied, and the signal was transmitted. An example from the receiver's perspective could be seen in Fig. \ref{fig:rx_fdom_example}. Controlling the center frequency in baseband allowed easier collection of metrics for OTA tests, since the frequency location of each signal was known by software.

\section{Spectrum Sensing Methodologies}
\label{sec:methodologies}

\subsection{Energy-based Spectrum Sensing}
\label{sec:energy_sensing}
\begin{figure}[htp]
    \centering
    \includegraphics[width=0.4\textwidth]{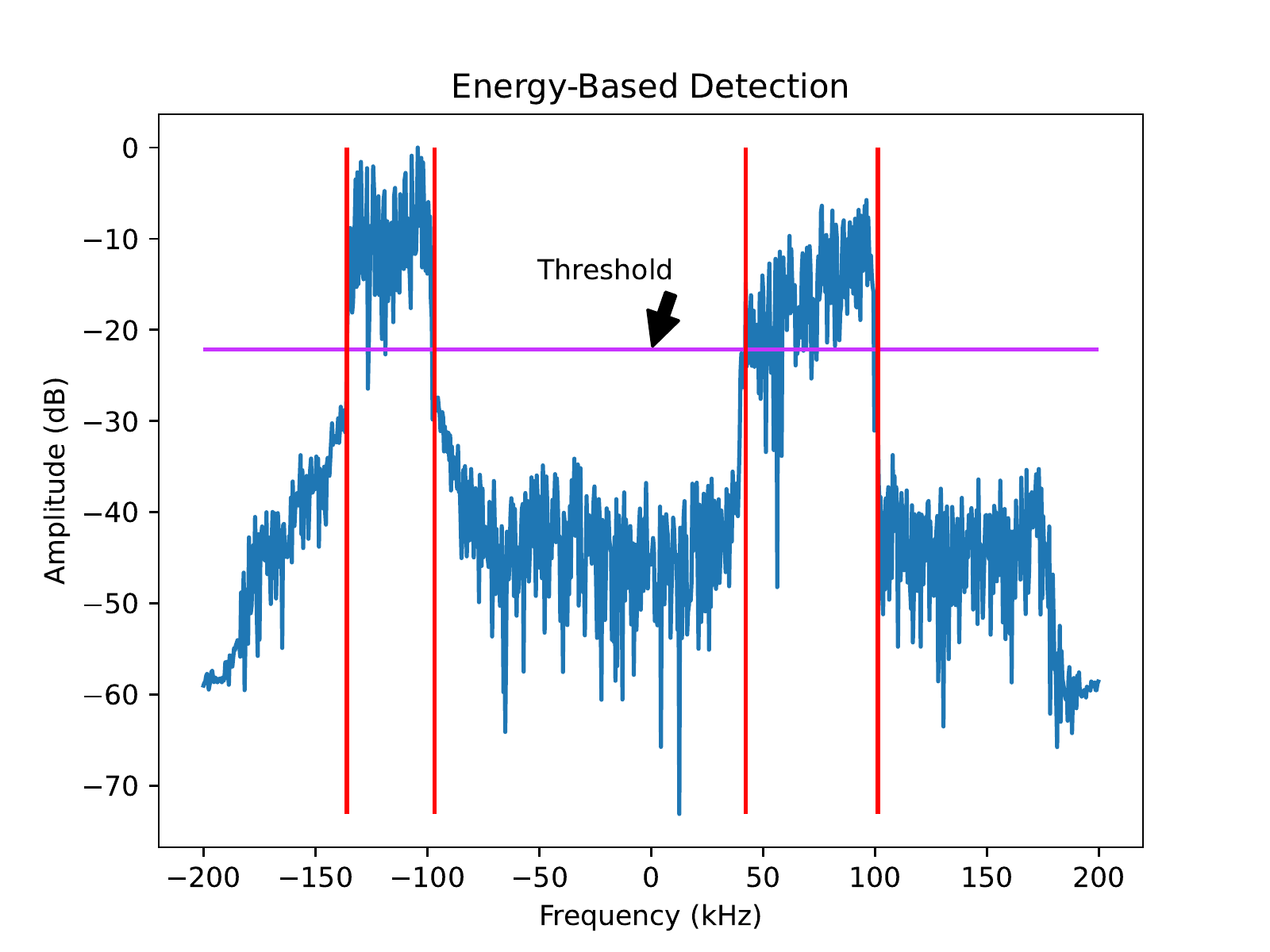}
    \caption{Example detection using energy-based spectrum sensing.}
    \label{fig:energy_example}
\end{figure}

To compare our results to popular energy-based spectrum sensing, we implement a base-line solution. This approach identifies when a transmitter is present when the signal energy exceeds a threshold. Multiple transmitters can be identified and localized using changepoint detection, which finds the points that exceed an energy threshold. In doing this, individual signals can be located. First, the FFT of a received signal is taken, so that its frequency contents can be observed. Next, the FFT output is scanned from left to right, looking for parts where the energy exceeds the noise floor. The intervals between these detections are considered a transmission, and the detections themselves are the frequency bounds. The noise floor must first be estimated, and a threshold determined from this. A simple estimate is to use bot the mean and standard deviation of a received signal frame, and to sum them together as $\gamma$. The equation for finding the noise floor threshold, $\gamma$, is shown in Eq. \ref{eq:noise_floor_thresh}, where $\mu$ is the mean of the FFT frame, $s_{i}$ is the $i$th sample of the frame, and $N$ is the length of the frame. For frames with a signal present, this will be slightly higher than the noise. For frames with no signal present, this will be the peaks of the noise. An example of this energy-based detection on an FFT frame could be seen in Fig. \ref{fig:energy_example}.

\begin{equation}
    \label{eq:noise_floor_thresh}
    \gamma = \mu + \sqrt{\frac{\sum^{N}(s_{i}-\mu)^2}{N}}
\end{equation}

\subsection{Spectrogram Spectrum Sensing using FRCNN}
\label{sec:2d_frcnn_sensing}
\begin{figure}[htp]
    \centering
    \includegraphics[width=0.4\textwidth]{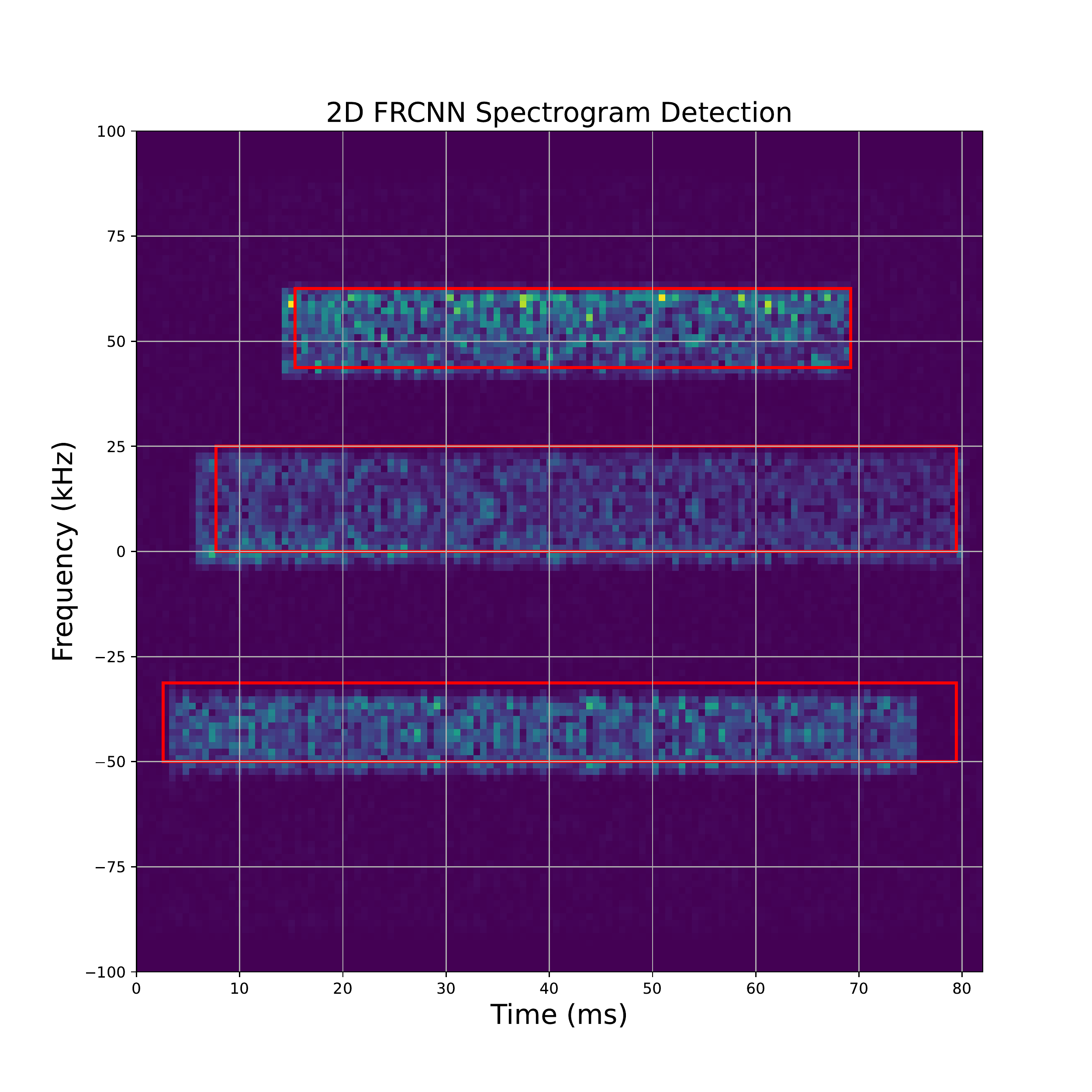}
    \caption{Example detection using spectrograms and 2D FRCNN.}
    \label{fig:spectrogram_example}
\end{figure}

The standard implementation of FRCNN can process 2D inputs, and therefor cannot be used to perform spectrum sensing on an FFT directly. To work around this, other authors have used the STFT to generate spectrograms, such as in Prasad et al \cite{prasad_deep_2020, prasad_downscaled_2020}. To compare our results with this approach, we implement 2D FRCNN spectrum sensing algorithm based on their approach. Since spectrograms are created by taking the FFT of many signals over time, they require significantly more data to create than if a single FFT was used. In order to produce a square image, the number of frames must be the FFT bin size. It is typical to use a smaller FFT size to reduce total data size, which comes at a tradeoff of reduced frequency resolution. We reduced the FFT size for spectrograms to 128, requiring 16,384 samples per spectrogram, compared to 1024 samples per frame for energy-based and 1D FRCNN spectrum sensing. Spectrograms incorporate time-frequency information into the same image. Therefor, an advantage of spectrograms is that the bounding boxes represent detections in both domains, unlike the energy-based and 1D FRCNN, which only detect frequency information. An example of using FRCNN on spectrograms could be seen in Fig. \ref{fig:spectrogram_example}.

\subsection{Spectrum Sensing using Optimized 1D FRCNN}
\label{sec:1d_frcnn_sensing}
\begin{figure}[htp]
    \centering
    \includegraphics[width=0.4\textwidth]{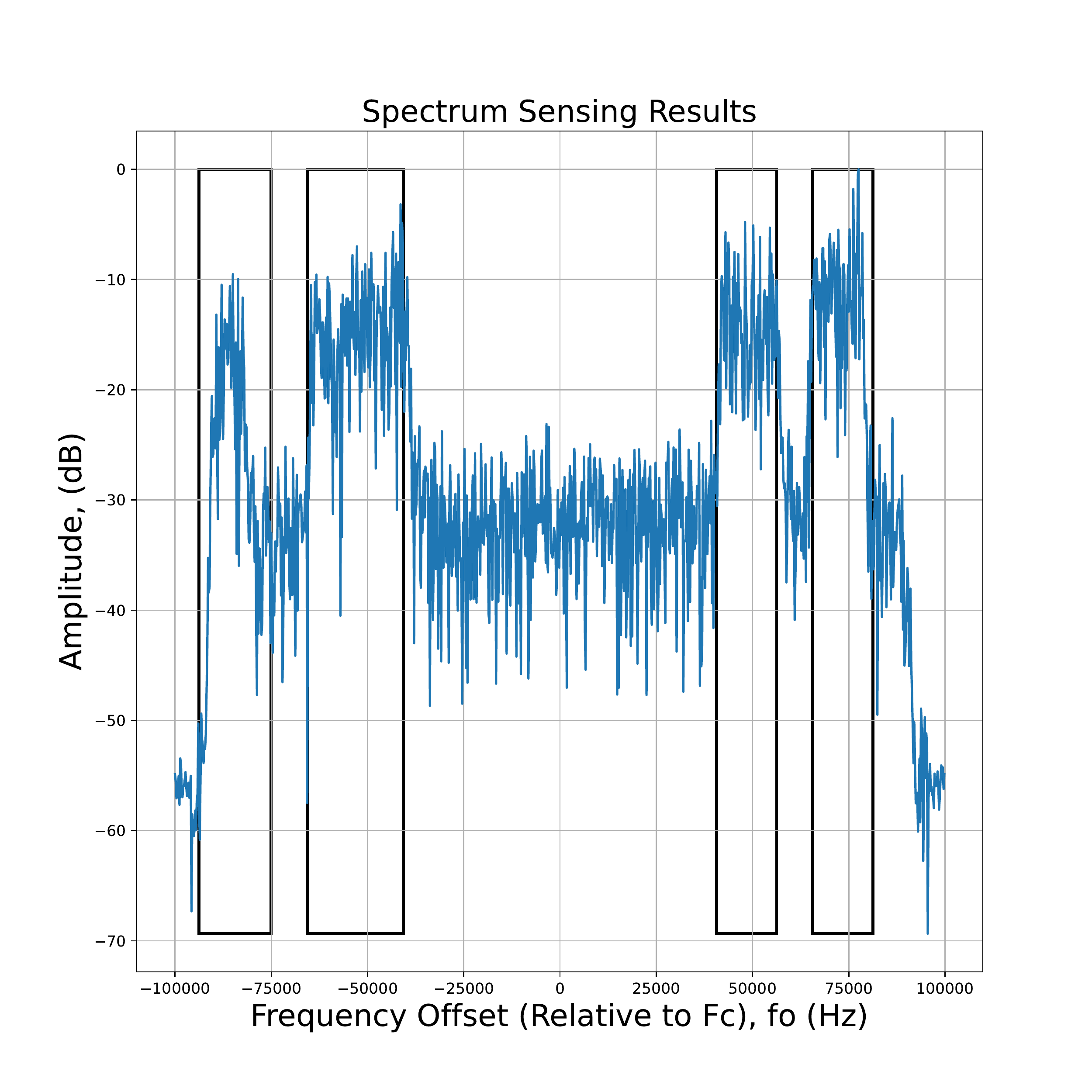}
    \caption{Example detection using FFTs and 1D FRCNN.}
    \label{fig:1d_example}
\end{figure}

In our optimized 1D FRCNN, we process FFT signals directly, bypassing the need to generate large spectrograms. The FFT is performed on a received baseband mixed-signal, and converted to decibels, which is a form that is easier to distinguish signals from noise. This resultant signal contains the same frequency information as a spectrogram equivalent, but consists of significantly less data compared to spectrograms. The resultant output from our optimized 1D FRCNN are bounding intervals, representing a single detected transmission, as well as the confidence in each prediction. These predictions are then thresholded by their confidence, such that only highly confident samples are considered positive detections. All other samples are discarded. An example of using 1D FRCNN on an FFT frame could be seen in Fig. \ref{fig:1d_example}.

\section{Measuring Performance}
\label{sec:metrics}
To measure performance, we use four metrics: The mean average precision (mAP), the mean intersection over union (mIoU), the probability of detection (Pd), and the probability of false alarm (Pfa). In object detection, the mAP is one of the most common metrics \cite{prasad_downscaled_2020, li_improved_2020, bassiouny_interpretable_2021, kim_attentive_2019, renu_chebrolu_deep_2019, wu_deep_2022, ren_faster_2015, liu_yolo-based_2022, prasad_deep_2020, vagollari_joint_2021, ghanney_radio_2020}. However, in spectrum sensing, the Pd and Pfa are the most common metrics for spectrum sensing \cite{xie_activity_2019, xie_deep_2020, zhao_eigenvalues-based_2021, liu_maximum_2019}. The mAP and mIoU are typical measurements for object detectors, while Pd and Pfa are typical metrics for spectrum sensing. In this section, we define these metrics, and how we measure them. These measurements may be different than other approaches, since we are comparing traditional energy-based methods to object detection methods. We consider the mIoU as our primary performance metric, as we are optimizing for the ability to localize all signals present, while minimizing false positive detections. However, other metrics give different insights to our model, that might be more relevant to other applications.

\subsection{Average Precision}
Object detection and data retrieval tasks often use the average precision (AP) or mean average precision (mAP) as performance measurements. The precision of a detector is the ability to make correct predictions, and recall is the ability to find all objects. Precision does not include missed objects, while recall does not include false predictions. To measure them, we first categorize predictions. Ground truths are objects that exist within an input. For example, the cars that exist within an image. Detections are the output object location and classifications of the detector. A true positive (TP) is when the detection overlaps a ground truth, where a false positive (FP) is a detection that overlaps no ground truth. A false negative (FN) is when there is no detection for a ground truth. True negatives (TN) are if there is no detection and no ground truth, and are not evaluated. We can count the number of TPs, FPs, and FNs by running the detector on a test dataset, and find the precision and recall by Eq. \ref{eq:precision} and Eq. \ref{eq:recall}. We can construct a precision-recall (PR) curve by measuring the precision and recall for each input image or signal. Finally, we can find the AP by taking the area under the PR curve.

\begin{equation}
    \label{eq:precision}
    \text{Precision} = \frac{\text{TP}}{\text{TP}+\text{FP}}
\end{equation}

\begin{equation}
    \label{eq:recall}
    \text{Recall} = \frac{\text{TP}}{\text{TP}+\text{FN}}
\end{equation}

To determine if a detection is true or false, we use the IoU, which measures the overlap between two objects, shown in Eq. \ref{eq:iou}. For 2D detection, these are typically bounding box detections, while for our 1D architecture, these are intervals. Typically, a detection is considered positive if the IoU with any ground truth is greater than 0.5, and negative if it is less than or equal to 0.5.

Detectors will give a probability for each detection. For FRCNN, the probability corresponds to a class. At minimum, there are two classes, foreground representing an object, and background representing no detection, however detectors can be configured to have more. We threshold the probability to determine if a detection is positive or negative. If the probability is above a threshold, the detection is considered positive. This can improve performance by removing poor samples. Since the probability threshold changes which detections are positive or negative, it also changes the precision and recall, and thus the AP. The mAP can measure a detector's performance across different thresholds, by finding the AP for each threshold, and averaging them together.

A common method of finding the AP is through interpolation. This method can be superior to the standard way of finding AP, since recall values may not be evenly distributed between [0,1], and may be concentrated around a single point. First, we sort each detection by its probability in descending order. Then, each prediction is determined to be a truth if its IoU with any ground truth is greater than 0.5, and false otherwise. The total number of false negatives are counted from this. Then, the precision and recall is calculated for each sample, with the number of true positives and false positives accumulating. Some recalls may have multiple precision values. For these, the maximum precision is used. Finally, the PR curve is divided into even spaced points, typically eleven. The precision at each point is averaged together to find the AP.

\subsection{Mean Intersection over Union}
In order to address some of the shortcomings from AP, we introduce the mean intersection over union (mIoU). The average precision tends to over estimate the ability of the detector in cases where the number of false positives are high, but false negatives are low. This is due to having mostly high recall, while precision can be very low. This is exacerbated by the common practice of assuming the PR curve starts at a precision of 1, and ends at a recall of 0, as well as in interpolated AP, where we take the maximum precision for each recall level. Additionally, the AP makes binary predictions of overlaps. If the IoU is greater than a threshold, then it is considered overlapping a ground truth. Therefor, the AP may overestimate how well the detector localizes objects. Finding the mean of IoUs can address both of these concerns. For spectrum sensing, many applications require accurate localization of transmissions in frequency domain. When sharing spectrum, inaccurate localization can result in poor spectrum usage, or interference between transmitters. Other sensing tasks may require learning information about a transmitter from its spectral patterns, or separating transmissions.

To find the mIoU, we first loop through every input signal and corresponding detection. For each detection, we calculate the IoU with all ground truths. If the IoU is greater than 0.5, it is considered overlapping, and the IoU is recorded. If the IoU is less than 0.5, or if it is a duplicate detection, the IoU is recorded as zero. Additionally, for any ground truths that had no detections overlapping, a value of zero is recorded for the IoU. For the multiclass case, misclassifications are considered a false positive, thus even if the IoU of a detection with a ground truth is high, it will still be considered zero if it is misclassified. Finally, the mean of all recorded IoUs is taken as the mIoU. An mIoU of zero is a detector that is unable to locate any objects, while an mIoU of one is a detector that locates all objects perfectly.

There are some shortcomings to the way we define and measure the mIoU. Both duplicate detections, and false positive detections over noise, are weighted equally. While these can both be viewed as negative, in some applications, duplicate detections may not be an issue. Additionally, detections that are close to overlapping, but do not have an IoU greater than 0.5, will be factored into the metric as an IoU of 0.0, which may underestimate the ability of the detector. An alternative to using the mIoU could be to find the AP over a range of IoU thresholds from 0.5 to just under 1.0, then averaging all APs together as the mAP. This would allow the mAP to incorporate the detector's ability to localize objects, but would not address the first issue, when false negatives are low and false positives are high. In this research, we focus on the mIoU as our primary metric, as it is the most fitting for spectrum sensing using object detection.

\subsection{Probability of Detection}
The probability of detection is defined as the probability that a signal is detected, $P_d=P(T > \gamma | H_{1})$, where $\gamma$ is an amplitude threshold and $H_{1}$ is the event that a signal exists. This is a common metric in spectrum sensing to measure a system's ability to find all signals present. In traditional spectrum sensing, this is the probability that the signal amplitude exceeds a threshold, given a transmitter is present. In the object detection case, this is better defined as the probability that a prediction overlaps a ground truth. To adapt this to the object detection case, we use the true positive rate (TPR), which is defined by Eq. \ref{eq:tpr}. Since FRCNN outputs bounding box predictions, they must first be sorted by true/false and positive/negative samples. To do this, we apply IoU and probability thresholding. Additionally, in the multiclass case, a misclassification will not be considered a detection. To keep results consistent, all models, including the energy-based spectrum sensing method, will evaluate their probability of detection with the TPR.

\begin{equation}
    \label{eq:tpr}
    \text{TPR} = \frac{\text{TP}}{\text{TP}+\text{FN}}
\end{equation}

\subsection{Probability of False Alarm}
The probability of false alarm is another common spectrum sensing metric, which is defined as the probability that a signal will be detected, when none exists, $P_{fa}=P(T > \gamma | H_{0})$, where $\gamma$ is a threshold, and $H_{0}$ is the event that no signal exists. In the object detection case, this is better defined as the probability that a detection does not overlap a ground truth. In other words, a false alarm is a detection over noise, or where the overlap with a signal is below a threshold. In the multiclass case, a misclassification will still be considered a false alarm. The false discovery rate (FDR) can be defined as the ratio of false positives to number of detections, shown in Eq. \ref{eq:fdr}. To keep results consistent, all models, including the energy-based spectrum sensing method, will evaluate their probability of false alarm with the FDR.

\begin{equation}
    \label{eq:fdr}
    \text{FDR} = \frac{n_{\text{fa}}}{n_{\text{det}}}
\end{equation}

\subsection{Inference Time}
Machine learning algorithms often take longer to process data than their signal processing counter parts. Downscaling FRCNN to process 1D signals reduces the amount of data that needs to be processed, and thus should decrease processing time. For ML/DL networks, the time it takes to run the model is called the inference time. In the FRCNN case, this is the time it takes to to find all signals present in an input, without including preprocessing times. To compare the different techniques, the average inference time is measured between energy-based, 1D FRCNN, and 2D FRCNN. For the FRCNN cases, this is the time between inputting an FFT or spectrogram into the network, and collecting each output detection, including postprocessing such as NMS. For energy-based, this is the time it takes to estimate an energy threshold, search the FFT, and organize all detections. The inference times are heavily dependent on hardware, and thus the absolute time measured may not be the most useful metric, thus when reporting inference time, we normalize the time relative to the baseline energy-based method.

\section{Analysis of FRCNN Parameters}
\label{sec:architecture_analysis}

The base of the network plays an important role in extracting features for the RPN and classifier. It directly controls key parameters such as resolution, inference time, memory usage, and complexity. Since the number of anchor boxes is directly related to the feature matrix size, increasing the stride or feature reduction will reduce the resolution of the RPN layer, and make each anchor represent a larger area of the input. However, increased number of layers can increase complexity, and improve the features that are extracted. Additionally, other parameters such as the number of filters in each layer, or the general structure of feature extraction layers, can impact the performance of the network. In this section, we evaluate the difference in performance for different configurations of the feature extraction layer. Then we conclude on a configuration that best suits the application at hand.

\subsection{VGG net feature extraction with different levels of stride}
\begin{figure}[htp]
    \centering
    \includegraphics[width=0.4\textwidth]{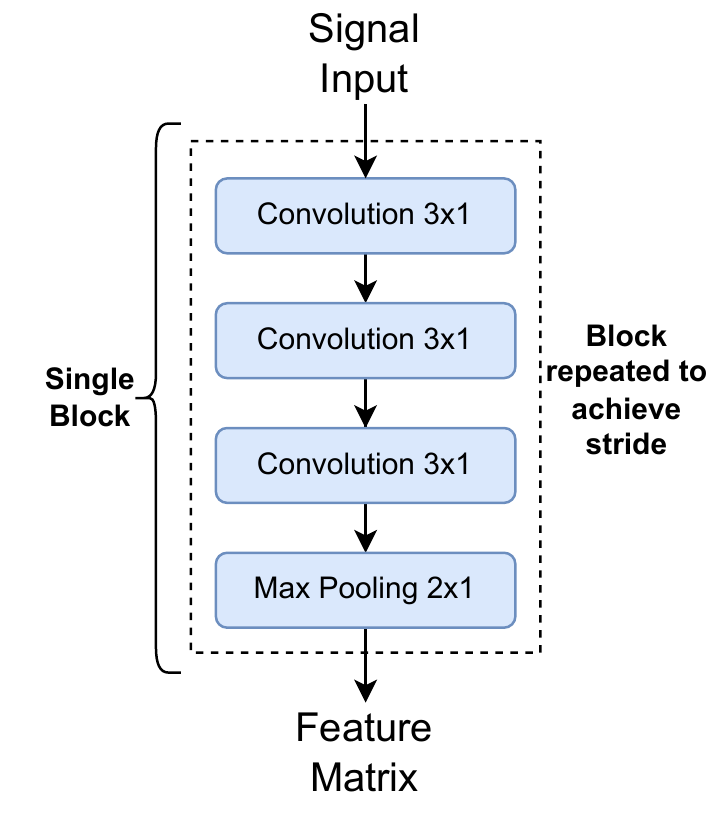}
    \caption{Architecture of the VGG feature extraction network.}
    \label{fig:vgg_example}
\end{figure}

The baseline architecture used is VGG net. In the original FRCNN paper \cite{ren_faster_2015}, the authors used VGG-16, pre-trained on ImageNet, as their feature extraction layers. Here, our baseline uses 19 layers instead of 16, achieving a stride of 16. An illustration of this architecture is shown in Fig. \ref{fig:vgg_example}. Each block consists of three convolutions, with a filter size of 3x1. To ensure the dimension of the feature matrix is proportional to the input signal, each convolution pads the output. Then, the block ends with a 2x1 max pooling layer, with 2x1 stride, dividing the input dimension by 2. The final block does not have a pooling layer. The number of blocks can be changed to achieve higher or lower stride, complexity, and more refined features. A lower stride results in less downscaling of the input, and thus more features. Since each feature represents an anchor box, the granularity of anchor boxes can be increased. In turn, a higher stride results in more downscaling of the input, and thus less features. However, these features will be more refined, include more of the input image, and reduce inference time. A trade-off must be struck between increasing the granularity of features, the number of anchor boxes, and the complexity. 

As a baseline, we use this VGG net setup. We test it with strides of 2, 4, 8, 16, and 32. Each increase in stride is an increment on the previous layers. These are summarized in Table \ref{tab:vgg_parameters}, as well as the number of filters for the output convolution.

\begin{table}[htb]
    \centering
    \begin{tabular}{|c|c|}
        \hline
        \textbf{Downscale Factor} & \textbf{Filter Count} \\
        \hline
        2 & 64 \\
        \hline
        4 & 128 \\
        \hline
        8 & 256 \\
        \hline
        16 & 512 \\
        \hline
        32 & 512 \\
        \hline
    \end{tabular}
    \caption{VGG Network Parameters}
    \label{tab:vgg_parameters}
\end{table}

\subsection{Resnet feature extraction}
\begin{figure}[htp]
    \centering
    \includegraphics[width=0.4\textwidth]{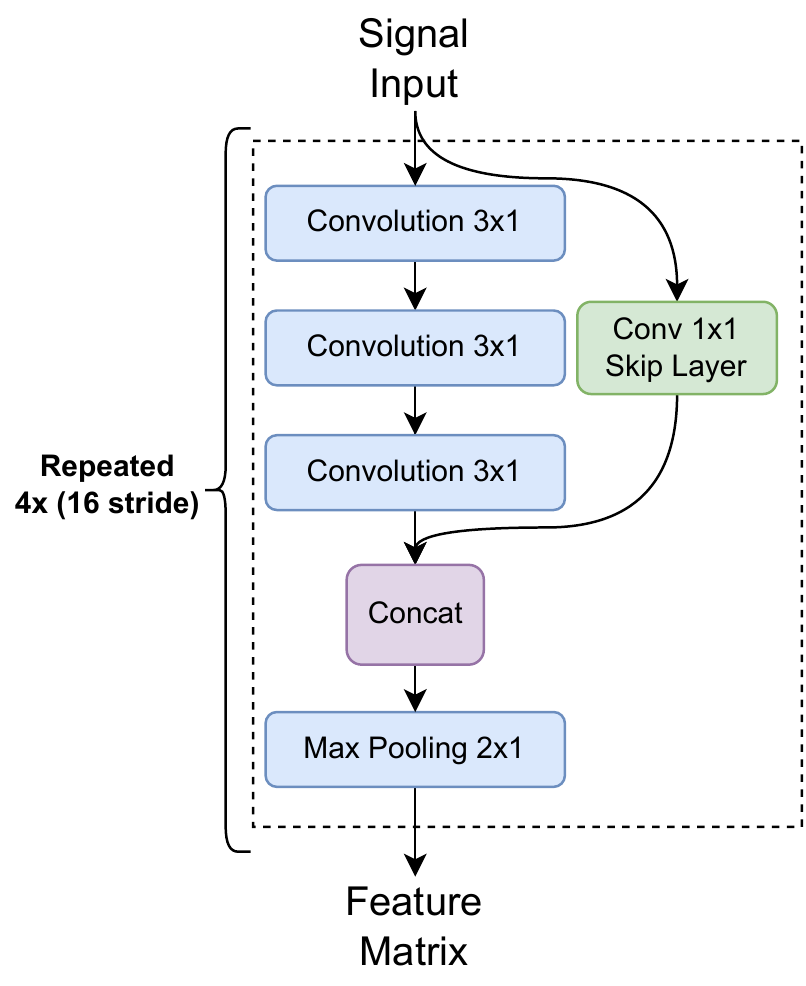}
    \caption{Architecture of the Resnet feature extraction network.}
    \label{fig:resnet_example}
\end{figure}

We test an additional common image processing architecture called Resnet, of which the typical architecture could be seen in Fig. \ref{fig:resnet_example}. The core of ResNet is the skip connection, which forwards parameters through the network. This allows some blocks to be skipped, if they do not contribute to the output. We compare the use of these skip layers, by applying them to the outlined VGG net. A skip layer is added in parallel with each block, where the features are concatenated together after the third convolution. Following this, batch normalization is added to improve training, and max pooling is performed to achieve data reduction. The resnet method was only tested for a stride of 16.

\subsection{Signal Processing Architecture}
\begin{figure}[htp]
    \centering
    \includegraphics[width=0.4\textwidth]{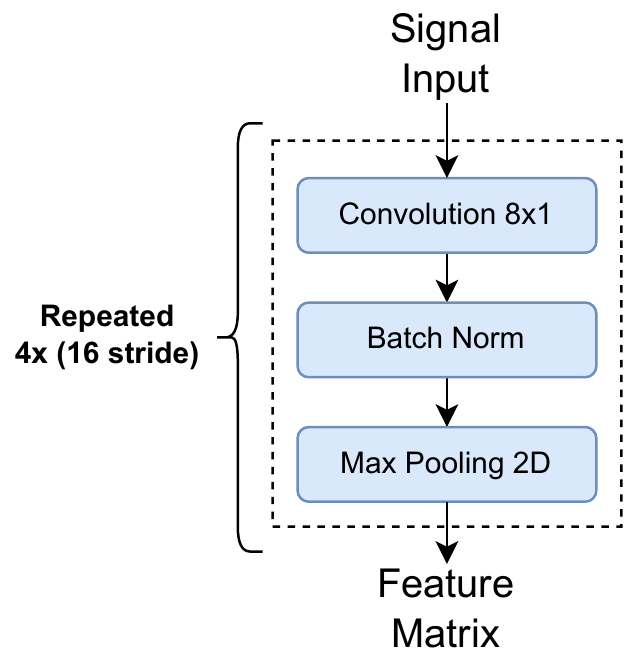}
    \caption{Architecture of the signal processing feature extraction network.}
    \label{fig:amc_network}
\end{figure}

In previous research, we used an architecture designed and tuned for signal processing \cite{morehouse_baseband_2020, morehouse_incremental_2021}, illustrated in Fig. \ref{fig:amc_network}. The network consists of linear blocks of convolutional layers, ReLU activation, batch normalization, and max pooling. The filter size of the convolutional layer was designed around the samples per symbol of the waveform, which relates the bandwidth of the signal to the receiver's sample rate. We test this architecture as well, using a stride of 16.

\subsection{Filter Sizes}
In Prasad et al \cite{prasad_downscaled_2020}, the authors downscaled FRCNN for the spectrum sensing case. While they still used images, they found that the high number of filters were unnecessary, and even detrimental, for this application. They reduced the number of filters in the RPN's first convolutional layer from 512 to 128. In the classifier, they reduced the number of filters in both fully connected layers from 4096 to 2048. We also test this downscaling for the VGG net with 16 stride, to see if we can achieve similar performance gains. In our results, we label this as the downscaled version.

\subsection{Comparison of Configurations}
\begin{figure}[htp]
    \centering
    \includegraphics[trim={5cm 8cm 5cm 8cm},width=0.4\textwidth]{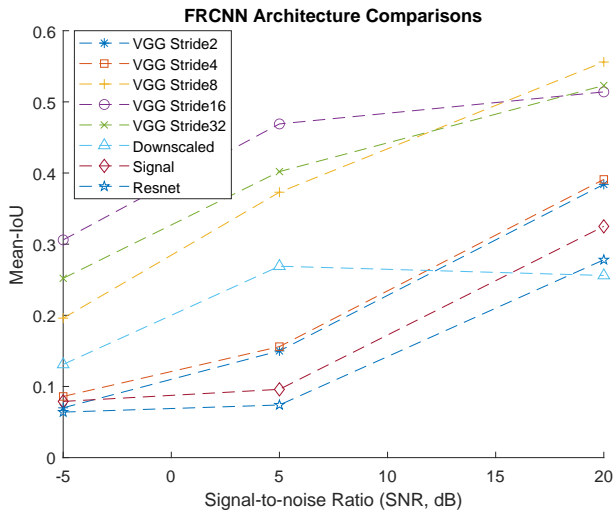}
    \caption{Mean IoU performance of different FRCNN architectures.}
    \label{fig:config_miou_results}
\end{figure}

\begin{figure}[htp]
    \centering
    \includegraphics[trim={5cm 8cm 5cm 8cm},width=0.4\textwidth]{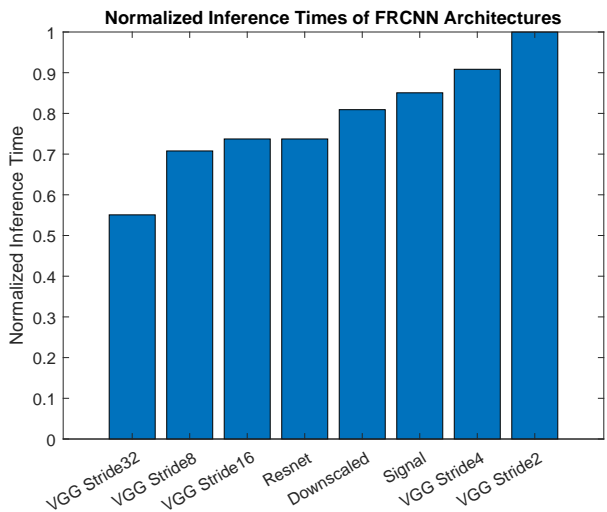}
    \caption{Inference times of different FRCNN architectures.}
    \label{fig:config_inference_results}
\end{figure}

To compare the architectures, each network was trained individually. Three datasets were generated, one for each training, validation, and testing. The same three datasets were used for training and evaluating each network. Each configuration was trained for epochs, with 10,000 samples each epoch. The Adam optimizer, with an initial learning rate of $10^-5$ was used. While higher performance could be achieved with more training epochs, a majority of learning is achieved in the first 4 epochs. This was verified using the validation loss. Thus, 4 epochs was chosen to reduce total training time to evaluate each architecture. We measure the mean-IoU at SNR levels of -5, 5, and 20 dB using synthesized data. We choose the mean-IoU here as it measures the detector's ability to: detect all signals present, to not raise false alarms, and localize each signal. These results are shown in Fig. \ref{fig:config_miou_results}. Additionally, we measure the average inference time for each configuration. This is the time it takes to process a single frame of data. We normalize the inference time to the highest measured, since absolute times will depend on the specific hardware used for testing. The comparison between different architectures inference time could be seen in Fig. \ref{fig:config_inference_results}. When comparing each architecture, we find that VGG net achieves the highest mIoU, with comparable inference times to other networks. VGG16 achieves 58\% higher mIoU than the second highest network, the signal network. Additionally, the inference time is approximately the same as Resnet, which are both the lowest, with 70\% relative inference time. Therefor, we choose VGG-net as our feature extraction network. Next, we evaluate the downscaling factor for our feature extraction network. As the downscaling factor increases, the inference time trends down, with the exception of factor of 8 being slightly faster than a factor of 16, although both are comparable. The highest mIoU is for VGG8, at 20dB SNR, but only by 6\% over a stride of VGG32. However, at lower SNR, VGG16 achieves 17\% higher mIoU over VGG32. Therefor, we choose VGG with a downscaling factor of 16 as our feature extraction layers due to its highest average mIoU, with decent inference time. However, it should be noted, that if inference time is more important, VGG32 is 25\% faster than VGG16, with comparable mIoU.

\section{A Use Case of 1D FRCNN in Signal Classification}
\label{sec:amc}
\begin{figure}[htp]
    \centering
    \includegraphics[width=0.25\textwidth]{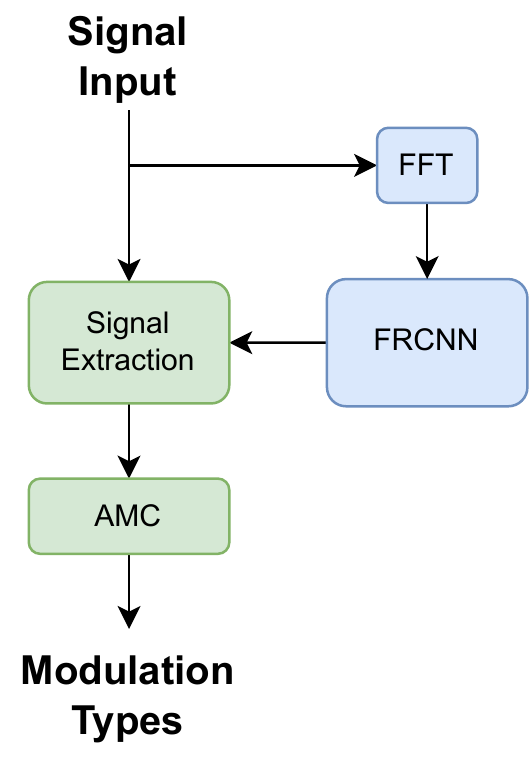}
    \caption{AMC Use Case Process.}
    \label{fig:use_case_flowchart}
\end{figure}

We provide a use case of our optimized 1D FRCNN using AMC, which is the process of identifying the modulation type of a given signal. In previous research, we used CNNs to achieve this with raw baseband samples \cite{gravelle_deep_2019, morehouse_baseband_2020}. In this use case, we show how this can be extended to the mixed signal case, to identify the modulation type of each transmitter present. The process is illustrated in Fig. \ref{fig:use_case_flowchart}. In this model, baseband samples from the receiver are supplied as the input. The model outputs the modulation type for every signal present. This process detects each transmitter in frequency domain, separates each transmission in time domain, then classifies the transmissions individually. First the FFT is performed on the received baseband samples, and the result is fed into the FRCNN model to find the location of all signals. Next, for each detection, the signal must be extracted from the baseband samples. This can achieved using the detected center frequency and bandwidth from FRCNN. This is done by removing the center frequency offset using Eq. \ref{eq:freq_shift}, where $s(t)$ is the input signal, and $f_{o}$ is the frequency offset. Next, a lowpass filter expressed in Eq. \ref{eq:lowpass_filter} is configured using the detected bandwidth, where $B$ is the bandwidth, $f_{s}$ is the sampling frequency, and $n$ is the nth tap of the filter. This is then applied by convolving the filter taps with the baseband samples. The result is the individual signal of interest, isolated from all other signals. Finally, the filtered baseband samples are fed into the AMC CNN, and a classification is produced. This same process could be applied to other signal processing applications, by replacing the modulation classification block with a different signal processing block.

\begin{equation}
    \label{eq:freq_shift}
    y(t) = s(t)\cdot e^{-j2\pi f_{o}t}
\end{equation}

\begin{equation}
    \label{eq:lowpass_filter}
    h[n] = \frac{B}{fs} \text{sinc}(\frac{B}{fs}\cdot n)
\end{equation}

\section{Results}
\label{sec:results}
We create four spectrum sensing methods including energy-based, 2D FRCNN with spectrograms, our optimized 1D FRCNN, and our 1D FRCNN with an AMC use case. For each of these methods, we measure the mAP, mIoU, probability of detection, and probability of false alarm. We first test each method offline, using the synthesized dataset, giving easy and accurate control over the SNR of the channel. Each metric is tested over the SNR range of -5dB to +20dB. Then, we test the optimized 1D FRCNN with AMC over-the-air, and measure the four metrics at a consistent SNR.

First, we create the baseline energy-based spectrum sensing method. We implement this in Python, using the numpy package. The implementation follows the process outlined in Sec. \ref{sec:energy_sensing}. For each frame, a noise threshold is calculated using Eq. \ref{eq:noise_floor_thresh}. Then, the FFT is searched to see where the amplitude exceeds the calculated threshold. To add hysteresis, the amplitude must drop below the threshold five consecutive times. This prevents sudden energy spikes from triggering the spectrum sensing algorithm. For each test SNR, a dataset is run independently at the specified SNR. Metrics are calculated after every image is processed. The energy-based process does not produce probabilities, therefor, all detections are assumed to have a probability of 1.0 for calculating metrics.

Next, a baseline 2D FRCNN spectrum sensing method using spectrograms is created. This implementation uses Yinghan Xu's \cite{xu_faster_2019} code as an implementation in Python using Tensorflow and Keras. The example is modified to support our spectrogram dataset, using uncompressed tiff files and no resizing of the spectrogram, as well as our changed hyperparameters and architecture, which are shown in Table \ref{tab:hyperparameters}. The architecture is trained over 40 epochs using 1000 spectrogram images. The dataset mixes spectrograms between 10dB and 40dB SNR. For testing, datasets at each SNR are created and tested independently. Each image is run individually, and checked against ground-truths. After an entire dataset for a specific SNR is run, the four metrics are calculated. A probability threshold of 0.7 is applied to each detection, where detections with probabilities below 0.7 are considered negative detections.

Finally, we test our optimized 1D FRCNN method. We use Yinghan Xu's \cite{xu_faster_2019} as a reference and baseline for FRCNN so that we could create our 1D optimized version. The hyperparameters used are shown in Table \ref{tab:hyperparameters}. The network is trained over 20 epochs, with each epoch containing 10,000 samples. We used 10,000 samples in the dataset to counteract overfitting that occurred at 2,000 samples. For testing, each SNR dataset is run and tested independently. After an entire dataset is run, the four metrics are calculated. A probability threshold of 0.9 is applied to each detection, where detections with probabilities below 0.9 are considered negative detections. This threshold was found to give the highest mIoU results after testing thresholds between 0.0 and 0.9.

\begin{table}[htb]
    \centering
    \begin{tabular}{|c|c|c|}
        \hline
        \textbf{Parameter} & \textbf{1D} & \textbf{2D} \\
        \hline
        rpn overlap min & 0.3 & 0.3\\
        \hline
        rpn overlap max & 0.7 & 0.7\\
        \hline
        nms overlap & 0.5 & 0.6\\
        \hline
        Optimizer & Adam & Adam \\
        \hline
        Learn Rate & $10^-5$ & $10^-5$ \\
        \hline
        Num Epochs & 20 & 40\\
        \hline
        Epoch Length & 10000 & 1000 \\
        
        \hline
    \end{tabular}
    \caption{Training Hyperparameters}
    \label{tab:hyperparameters}
\end{table}

\begin{figure}[htp]
    \centering
    \includegraphics[width=0.4\textwidth,trim={4cm 8.5cm 4cm 9cm}]{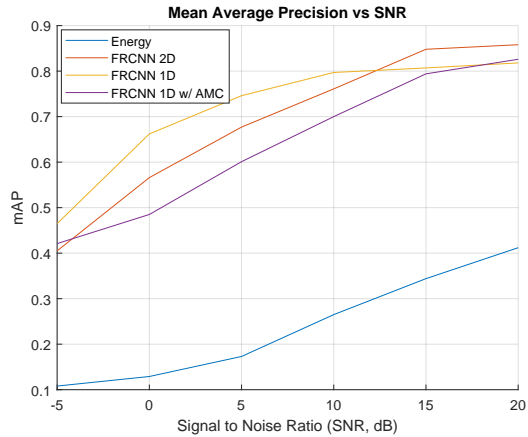}
    \caption{Mean Average Precision vs SNR}
    \label{fig:map_vs_snr}
\end{figure}

\begin{figure}[htp]
    \centering
    \includegraphics[width=0.4\textwidth,trim={4cm 8.5cm 4cm 9cm}]{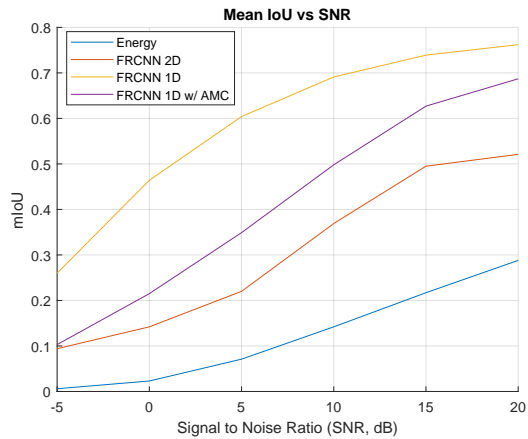}
    \caption{Mean Intersection over Union vs SNR}
    \label{fig:miou_vs_snr}
\end{figure}

\begin{figure}[htp]
    \centering
    \includegraphics[width=0.4\textwidth,trim={4cm 8.5cm 4cm 9cm}]{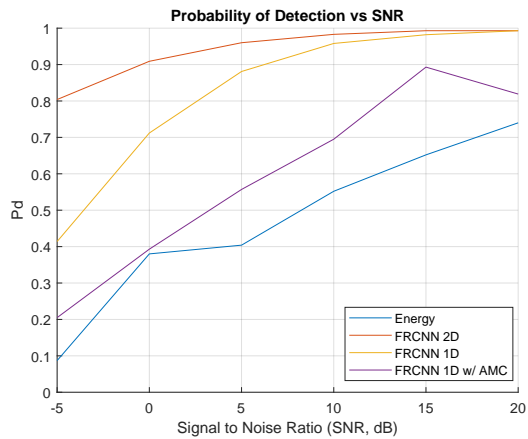}
    \caption{Probability of Detection vs SNR}
    \label{fig:pd_vs_snr}
\end{figure}

\begin{figure}[htp]
    \centering
    \includegraphics[width=0.4\textwidth,trim={4cm 8.5cm 4cm 9cm}]{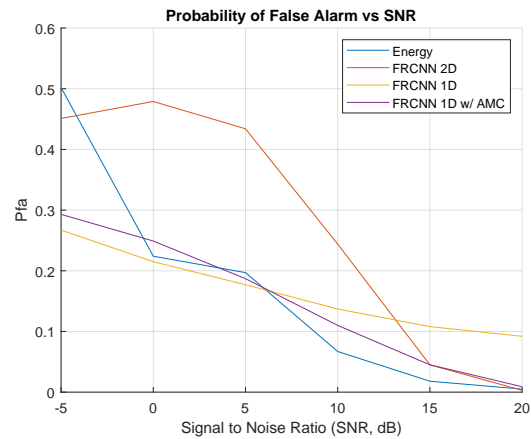}
    \caption{Probability of False Alarm vs SNR}
    \label{fig:pfa_vs_snr}
\end{figure}

The mAP measured over SNR could be seen in Fig. \ref{fig:map_vs_snr}. The average mAP over SNR is 0.239 for energy-based, 0.686 for 2D FRCNN, and 0.716 for our 1D FRCNN spectrum sensing. Both FRCNN methods significantly outperform energy-based sensing. Our method outperforms 2D spectrograms at low SNR, however, the mAP falls below 2D after 10dB SNR. The higher mAP shows the ability of our method to correctly detect all signals, without making false detections. The mIoU measured over SNR could be seen in Fig. \ref{fig:miou_vs_snr}. The average mIoU over SNR is 0.125 for energy-based, 0.307 for 2D FRCNN, and 0.587 for our 1D FRCNN spectrum sensing. Under all SNR conditions, our method localizes transmissions better than other approaches. The Pd measured over SNR could be seen in Fig. \ref{fig:pd_vs_snr}. The average Pd over SNR is 0.469 for energy-based, 0.940 for 2D FRCNN, and 0.823 for our 1D FRCNN spectrum sensing. The 2D spectrogram method has the highest probability of detecting transmissions, especially at low SNR. Our 1D method had higher Pd than energy-based at low SNR, but significantly under-performed compared to 2D, however, as SNR increases, both 1D and 2D methods converged to a Pd of nearly 1.0. The Pfa measured over SNR could be seen in in Fig. \ref{fig:pfa_vs_snr}. The average Pfa over SNR is 0.169 for energy-based, 0.276 for 2D FRCNN, and 0.166 for our 1D FRCNN. Our optimized version performs the best on average, over all SNRs. However, energy-based averages almost the same, and performs significantly better at high SNR. The 2D spectrogram method performs poorly at low SNR, but the Pfa quickly reaches the same levels as other methods at 15dB SNR. The relative inference time of each method is shown in Table \ref{tab:inference_times}. These are normalized to the highest inference time, 2D FRCNN. The energy-based method significantly outperformed both ML-based methods, being about 500 times faster than our optimized 1D method. However, our optimization to FRCNN significantly improved inference time, by a factor of 4 over the original 2D FRCNN.

\begin{table}[htb]
    \centering
    \begin{tabular}{|c|c|}
        \hline
        \textbf{Method} & \textbf{Relative Time}\\
        \hline
        Energy-Based & $5.00\cdot10^-4$ \\
        \hline
        1D FRCNN & $0.250$\\
        \hline
        2D FRCNN & $1.00$ \\
        \hline
    \end{tabular}
    \caption{Normalized Inference Times}
    \label{tab:inference_times}
\end{table}

Our optimizations to FRCNN significantly improve the localization performance and the inference speed for spectrum sensing. The mIoU, our primary performance metric, increases by 91\%, however the mAP increases only slightly by 4\%. Compared to energy-based, both FRCNN methods significantly outperform at all SNRs. Our optimized version improved mIoU by 470\%, and mAP by 300\% compared to energy-based sensing. For detecting all transmissions present, 2D FRCNN performed the best at low SNR, while both our optimized version and 2D perform just as well at SNRs above 15dB. Under the SNR intervals tested, energy-based under-performs compared to FRCNN methods, however, it trends towards the same Pd for increasing SNR. Finally, when comparing probability of false alarm, our method had a low Pfa at very low SNR, however, the false alarm rate only marginally improved as SNR increased. Meanwhile, both energy-based and 2D FRCNN had significantly lower Pfa at higher SNR. 2D FRCNN had high Pfa at low SNR, being almost 2x higher than the other methods at 5dB SNR. Our optimized FRCNN shows improved detection and localization performance, as well as faster inference time over 2D FRCNN. In applications where detecting all signals present is prioritized, 2D FRCNN still outperforms our optimized version. Additionally, only the spectrogram case with 2D FRCNN can jointly localize in time and frequency domain, if an application requires that. Energy-based still fits best in applications that require fast inference times, significantly outperforming both ML based methods.

\subsection{AMC Use Case Results}
We test our AMC use case using the same outlined metrics, however the dataset must change to include mixed-signal time-domain samples, as well as the class labels for each detection. The time-domain samples are used after FRCNN detection to isolate each signal, so that AMC can be performed on it. The class labels refer to the modulation of each time-domain signal. First, a dataset was generated to train the AMC CNN described in Sec. \ref{sec:amc}, created from 9,000 samples of the 1D FRCNN dataset. Each sample of the FRCNN dataset contained on average three transmissions, with random modulation types. Each transmission was lowpass filtered to isolate it, and the frequency offset was removed. To simulate errors in FRCNN, we apply a $\pm$2kHz random frequency offset, and a bandwidth offset between 70\% and 150\% of the signal's original bandwidth. Additionally, we window an empty part of the spectrum to simulate a ``No Signal'' case. The AMC model is then trained over 40 epochs, with a total of 37548 training samples, 2492 validation samples, and 12533 test samples. The large number of test samples was chosen to ensure overfitting was not occurring. After training, the combined FRCNN and AMC system was tested, and metrics were measured for mAP, mIoU, Pd, and Pfa over the SNR range of -5dB to +20dB. These results could be seen compared to energy based, 2D FRCNN, and our optimized 1D FRCNN without AMC, in Fig. \ref{fig:map_vs_snr}-\ref{fig:pfa_vs_snr}. First, it should be noted that compared to standalone FRCNN, combining AMC reduces the mAP, mIoU, and Pd. The ability of the system to detect and localize signals does not change, however, mislabelling signals will reduce each of these metrics. For example, a detected signal classified as ``BPSK'', when it was actually ``PAM4'', would factor in as a false positive detection, and an IoU of 0. The average mAP drops by 11\% to 0.638, the average mIoU drops by 30\% to 0.413, and the Pd drops by 28\% to 0.594. However, the Pfa improves, dropping by 10\% to 0.149, significantly outperforming all models. The Pfa metric is not effected by false positives over ground truths, and thus misclassifications, but the introduction of a ``No Signal'' class filters out additional false positive detections over noise. This use case shows that our optimized FRCNN model can be applied to mixed signal classification, when multiple and unknown signals are present in an uncontrolled band.

\subsection{Over-the-Air Results}
\begin{figure}[htp]
    \centering
    \includegraphics[width=0.4\textwidth]{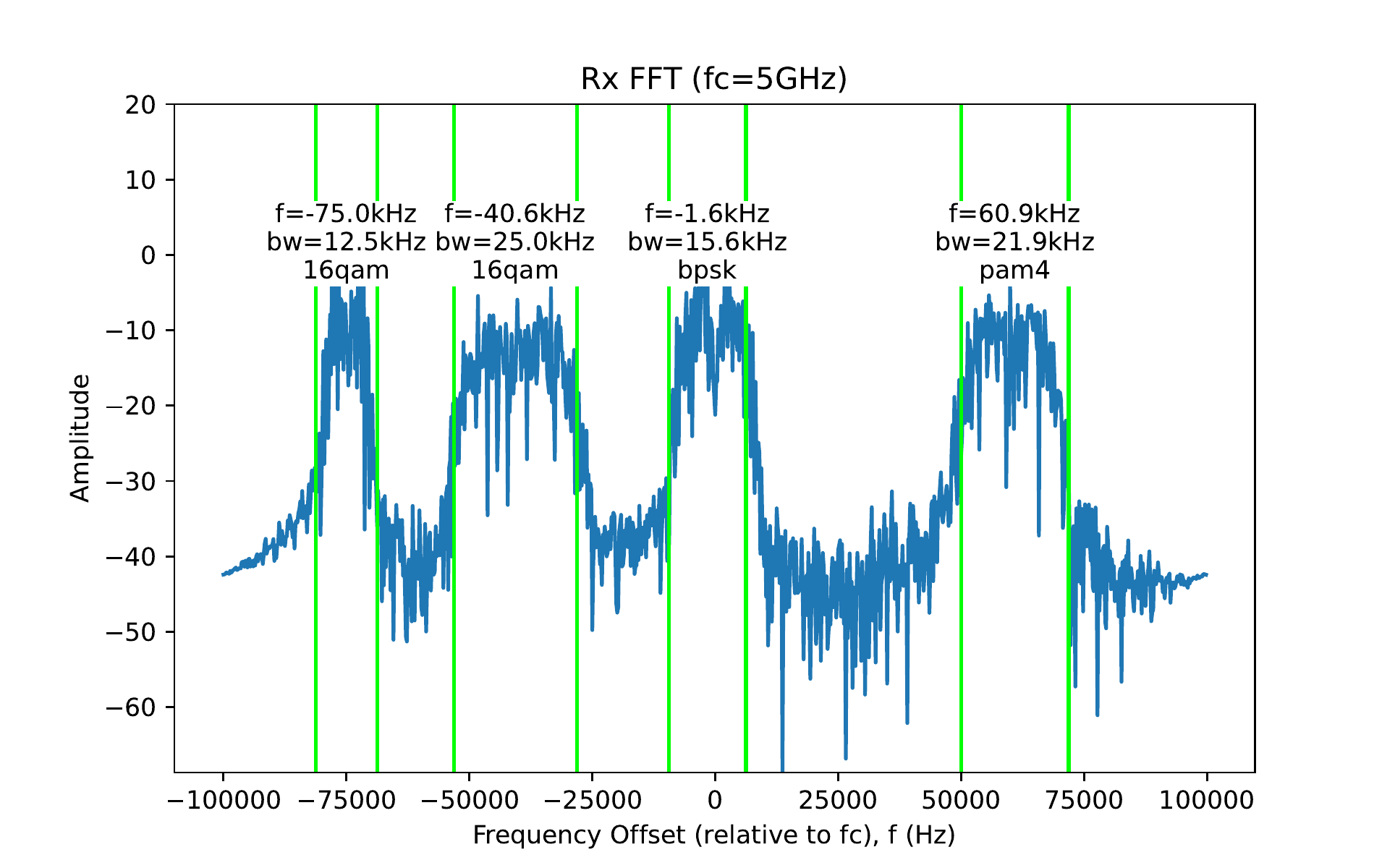}
    \caption{OTA Run 1 with FRCNN Detection.}
    \label{fig:ota_run_1}
\end{figure}

\begin{figure}[htp]
    \centering
    \includegraphics[width=0.4\textwidth]{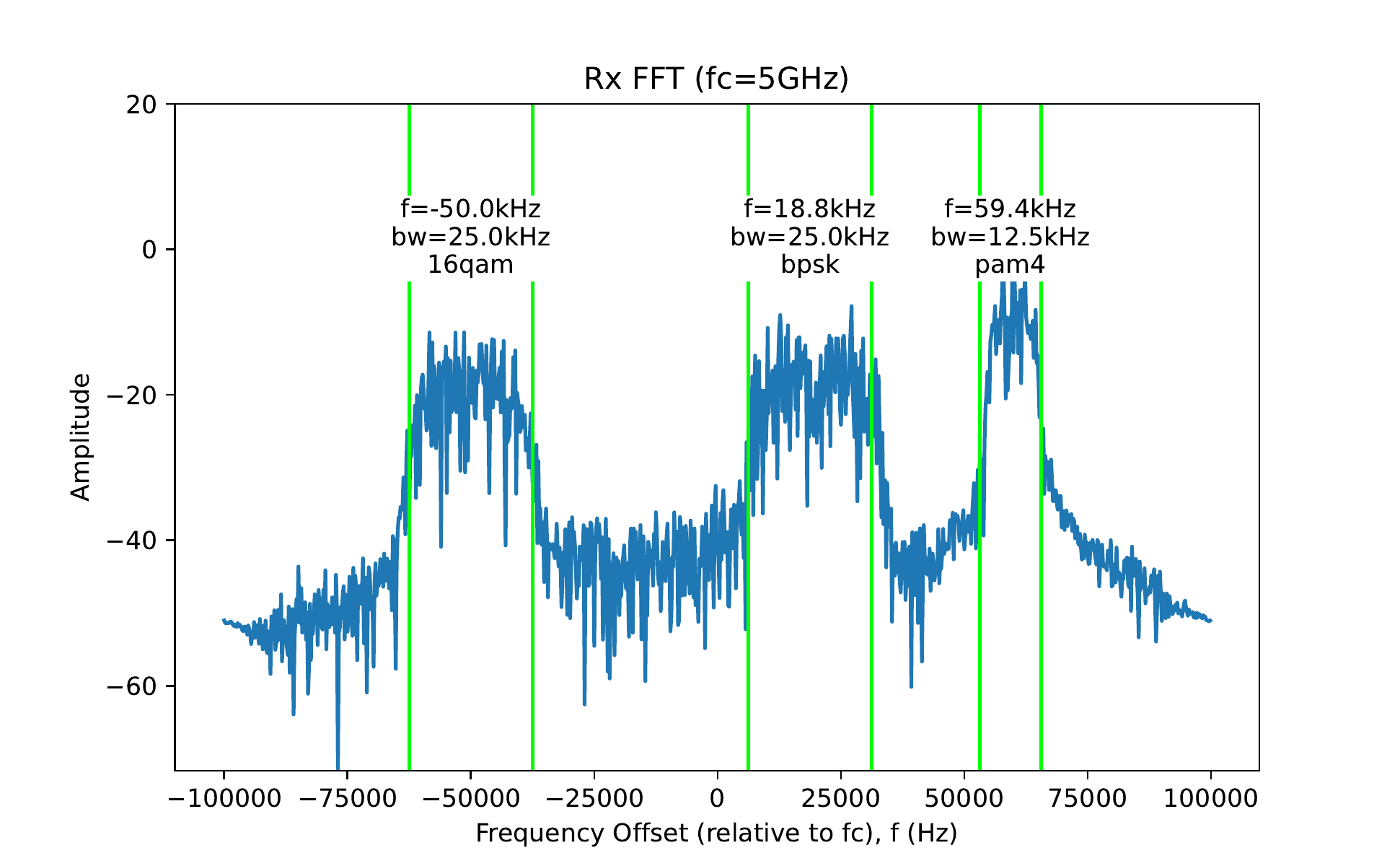}
    \caption{OTA Run 2 with FRCNN Detection.}
    \label{fig:ota_run_2}
\end{figure}

\begin{figure}[htp]
    \centering
    \includegraphics[width=0.4\textwidth]{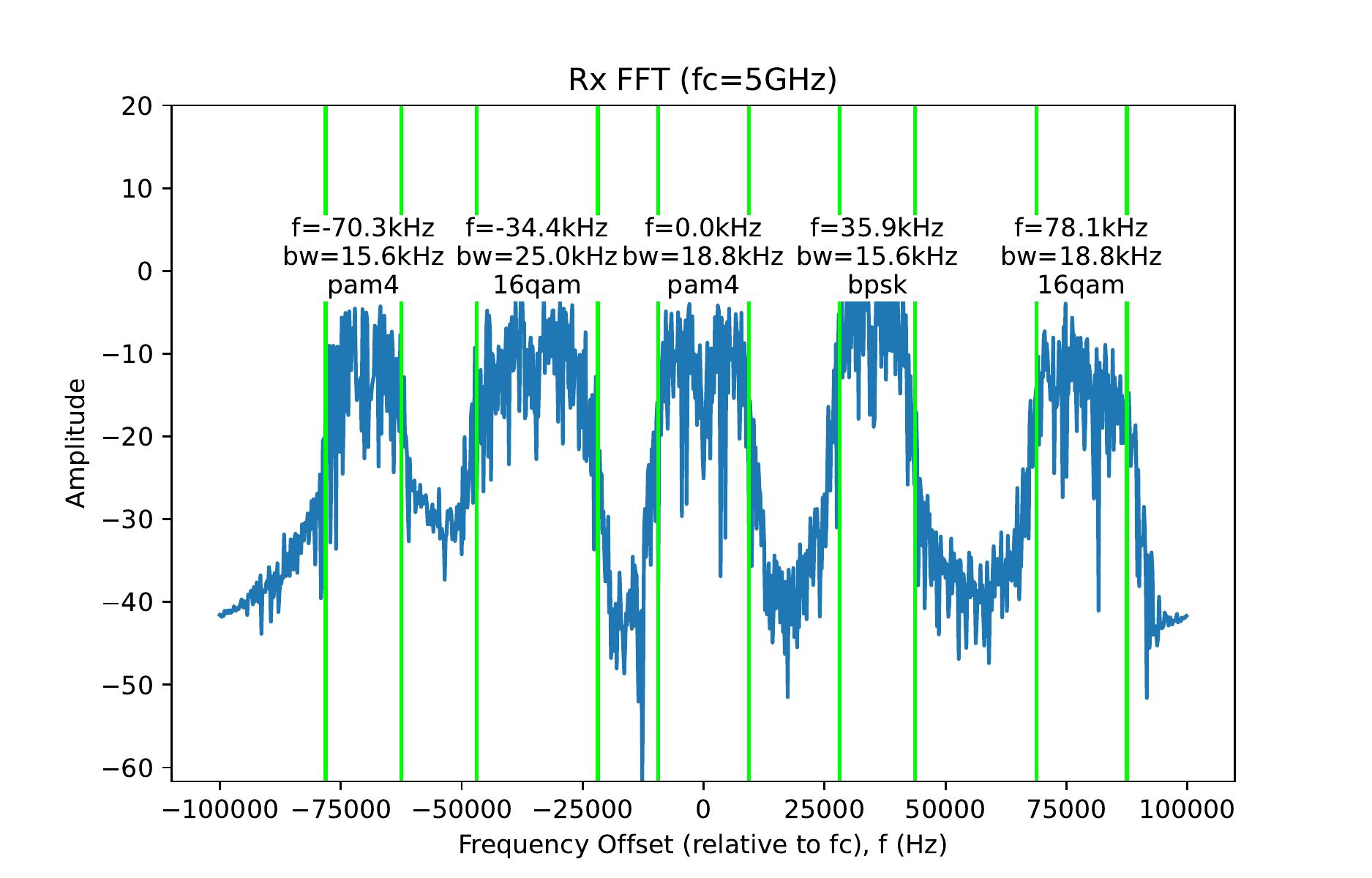}
    \caption{OTA Run 3 with FRCNN Detection.}
    \label{fig:ota_run_3}
\end{figure}

Finally, we test our system OTA by using SDR. We test our combined AMC and FRCNN model, to show the feasibility of this system to a realistic and real-world environment. We used the same trained models for 1D FRCNN and AMC that we collected metrics on. Two USRP N2901 SDRs were used to create the cluttered environment, where a total of five simultaneous transmissions were tested. A single antenna receiver observed the spectrum, and applied our spectrum sensing and AMC algorithm. Each SDR channel was configured to operate at a 5GHz center frequency, with a 200kHz baseband sample rate. GNURadio was used to interface the SDRs with a PC, and ZMQ was used to transfer samples between GNURadio and Python for ML. A total of three different transmitter configurations were run, described in Table \ref{tab:ota_configs}. These were created by generating the transmit waveforms in Python, and frequency multiplexing them in baseband. When received, this has the same effect as transmitting each signal at a different center frequency. Received samples were binned into frames of 1024, and processed using FRCNN. Each run was conducted until a total of 1000 frames were received, and each frame was processed online, as they were received. Since the SDR frame period was approximately 5ms, while the FRCNN inference time on our hardware was approximately 500ms, the system could not be run in real time. Instead, the spectrum was periodically observed, at the same rate as the inference time. While FRCNN was processing the received samples, current received samples were dropped, to ensure only the most recent samples were used. An FFT plot of each run could be seen in Fig. \ref{fig:ota_run_1}-\ref{fig:ota_run_3}, where the estimated center frequency, bandwidth, and modulation type are displayed above each detection. After running each test, the mAP, mIoU, Pd, and Pfa were calculated for each run, then averaged together, and could be seen in Table \ref{tab:ota_results}. Compared to offline tests at 20dB SNR, the mAP, mIoU, and Pd remained the same, with at most a 1.6\% change between mAP. The Pfa decreased significantly, by about 90\%, from 0.009 to 0.001. This is most likely due to the OTA SNR being higher. Compared to other literature that measures OTA performance, we show a significant improvement in OTA results. In Prasad et al \cite{prasad_downscaled_2020}, the authors found that their model did not generalize OTA, and produced an mAP of only 0.125. With our mAP of 0.826, we show that our system is capable of generalizing to real-world environments, and performing significantly better OTA.

\begin{table}[htb]
    \centering
    \begin{tabular}{|c|c|c|c|}
        \hline
        \textbf{Signal Number} & \textbf{$f_{c}$} & \textbf{$B$} & \textbf{Mod Type} \\
        \hline
        \multicolumn{4}{|c|}{Run 1} \\
        \hline
        
        \hline
    \end{tabular}
    \caption{OTA Configurations}
    \label{tab:ota_configs}
\end{table}

\begin{table}[htb]
    \centering
    \begin{tabular}{|c|c|c|c|c|}
        \hline
        \textbf{} & \textbf{Run 1} & \textbf{Run 2} & \textbf{Run 3} & \textbf{Average} \\
        \hline
        \textbf{mAP} & 0.728 & 0.885 & 0.904 & 0.839 \\
        \hline
        \textbf{mIoU} & 0.509 & 0.739 & 0.817 & 0.688 \\
        \hline
        \textbf{Pd} & 0.675 & 0.877 & 0.932 & 0.828 \\
        \hline
        \textbf{Pfa} & 0.000 & 0.002 & 0.000 & 0.001 \\
        \hline
    \end{tabular}
    \caption{OTA Results}
    \label{tab:ota_results}
\end{table}


\section{Conclusion}
\label{sec:conclusion}
In this research, we optimized FRCNN for 1D spectrum sensing, enabling FRCNN to be applied to 1D signals, a feature that was not previously possible. To perform spectrum sensing, we preprocess received signals with the FFT, to reveal frequency information. RF data was synthesized to train the optimized algorithm for spectrum sensing, as well as for testing the performance. Our optimizations improved the model in localization performance and inference time. Overall, the mIoU averaged 470\% improvement over energy-based, and 91\% improvement over 2D spectrograms. Optimizing for the 1D case improved inference time by 4 times compared to 2D spectrograms, however, compared to energy-based detection, ML approaches are still significantly slower, even with our optimizations. While our model performed well in detecting transmissions, 2D spectrograms had higher Pd at low SNR, while energy based had lower Pfa at high SNR.
To demonstrate the application of our optimizations, we provide a use case for multi-signal AMC. Time-domain samples are filtered to isolate individual signals based on FRCNN detections. Since this introduces a new source of error, labelling error, it causes the mAP and mIoU to drop slightly. However, by also analyzing signals in time domain, false negatives can be filtered out, greatly improving the false alarm rate. The Pfa for our AMC use case is comparable to energy-based.
Over-the-air tests were performed with our optimized FRCNN and AMC use case. We used two USRP N2901, to transmit up to 5 signals simultaneously, within a 200kHz band. We demonstrate that a single antenna receiver can use our architecture to accurately locate all signals. The demonstration was performed online, to show the ability to integrate the model with a receiver system. Compared to other papers that collected OTA metrics, our model performed significantly better. In Prasad et al \cite{prasad_downscaled_2020} the authors measured an OTA mAP of 0.125, while our mAP was measured 0.826. However, our work could still benefit from several improvements, including faster inference times, improved probabiltiy of detection and probality of false alarm, and greater performance at low SNR. In future applications, we plan to extend our work to other object detection in signal processing, integration into larger systems, and implementation on an FPGA for real-time acceleration. Our results contribute to broader spectrum sensing research, as well as object detection in signal processing. We show improved spectrum sensing results, beyond state of the art, for cluttered RF environments. Additionally, we improve the inference time over previous FRCNN methods. Our work can also be extended to other signal processing applications, where object detection is useful, such as localizing anomalies in a signal.


\section*{Acknowledgements}

This work was supported by AFRL Beyond 5G SDR University Challenge program, the University of Massachusetts Dartmouth’s Marine and Undersea Technology (MUST) Research Program funded by the Office of Naval Research (ONR) under Grant No. N00014-20-1-2170, and ONR Naval Engineering Education Consortium (NEEC) program under Grant No. N00174-22-1-0008. 

\bibliographystyle{./IEEEtran}
\bibliography{./IEEEabrv,./IEEEexample}

\end{document}